\documentclass[10pt,twocolumn,preprintnumbers,amsmath,amssymb,nofootinbib,superscriptaddress,prd]{revtex4-2}

\usepackage[dvipsnames]{xcolor}
\usepackage{bm}
\usepackage{graphicx}
\usepackage{tabularx}
\usepackage[colorlinks,allcolors=RoyalBlue]{hyperref}
\usepackage[normalem]{ulem}
\newcommand{\mnu}{M_\nu}

\newcommand{\lowLEE}{low-$\ell$ EE}

\newcommand{\chisqnoEE}{(\chi^2-\chi^2_{\tau})}

\begin{document}
\title{Turning a negative neutrino mass into a positive optical depth}

\author{Tanisha Jhaveri}
\affiliation{Kavli Institute for Cosmological Physics, Enrico Fermi Institute, and Department of Astronomy \& Astrophysics, University of Chicago, Chicago IL 60637
}

\author{Tanvi Karwal}
\affiliation{Kavli Institute for Cosmological Physics, Enrico Fermi Institute, and Department of Astronomy \& Astrophysics, University of Chicago, Chicago IL 60637
}

\author{Wayne Hu}
\affiliation{Kavli Institute for Cosmological Physics, Enrico Fermi Institute, and Department of Astronomy \& Astrophysics, University of Chicago, Chicago IL 60637
}

\begin{abstract}
Under $\Lambda$CDM, recent baryon acoustic oscillation (BAO) distance measures from DESI, which favor a low matter density $\Omega_m$, are in moderate $2-3\sigma$ tension with cosmic microwave background (CMB) observations.  This tension appears alternately as a preference for the sum of neutrino masses dropping below the $\sum m_\nu =  0.06$\,eV value required by neutrino oscillation measurements to formally negative values; a discrepant value of $\Omega_m$ at 0.06\,eV; or preference for dynamical dark energy beyond $\Lambda$CDM.  We show that this tension largely arises from the CMB lensing constraints on the calibration of the sound horizon for geometric measurements and relies on the measurement of the reionization optical depth $\tau$ from large-angle CMB polarization to set the lensing amplitude.   Dropping these constraints removes the neutrino tension at $\sum m_\nu=0.06$\, eV entirely, favoring $\tau = 0.091\pm 0.011$ in $\Lambda$CDM.  Beyond $\Lambda$CDM, it brings the preference for $w_0-w_a$ dynamical dark energy to below $95\%$ CL.   We explore the freedom in interpreting the \lowLEE\ polarization constraint due to analysis choices and reionization modeling beyond the standard step-function assumption and find that this drops the neutrino tension in $\Lambda$CDM to below $95\%$ CL.   Alternately, this raising of $\tau$ can also be achieved by the same reduction in large-scale curvature fluctuations that also ameliorates the low-$\ell$ temperature anomaly.
\end{abstract}

\date{\today}

\maketitle

\section{Introduction} 

Within the highly successful $\Lambda$CDM cosmological paradigm, the sum of the  neutrino masses $\mnu\equiv \sum m_\nu/\mathrm{eV}$ can be measured by its cosmological influence on the expansion history and the suppression of the growth of structure below the neutrino free-streaming scale (see\ \cite{ParticleDataGroup:2024cfk,DiValentino:2024xsv} for recent reviews).
On the other hand, neutrino-oscillation experiments precisely measure the splittings in the squared masses and consequently constrain the minimum $\mnu$. For normal ordering, where the smaller solar-oscillation splitting is between the lightest states, this minimum is $\mnu\approx 0.06$, and for the inverted ordering where it is between the heaviest states, the minimum is $\mnu\approx  0.1$ \cite{Esteban:2024eli}.

Recent measurements of the expansion history by 
the Dark Energy Spectroscopic Instrument (DESI) \cite{DESI:2024mwx} using baryon acoustic oscillations (BAO), when combined with cosmic microwave background (CMB) measurements, have called into question the compatibility between these minimal masses and $\Lambda$CDM \cite{Craig:2024tky,Green:2024xbb,Loverde:2024nfi,DESI:2025ejh,Lynch:2025ine,Naredo-Tuero:2024sgf,RoyChoudhury:2024wri, Reboucas:2024smm,Jiang:2024viw}.
The posterior probability of $\mnu$ peaks at zero and if formally continued, would favor unphysically negative values. With the recent second data release of DESI, this unphysical continuation gives
$\mnu{}_{\rm ,eff}=-0.101^{+0.047}_{-0.056}$ when combined with the CMB \cite{DESI:2025ejh}, which implies a fairly significant tension with even the normal ordering bound under $\Lambda$CDM. This tension can alternately be viewed as supporting a dynamical dark energy extension of the paradigm. 

In fact, the origin of the CMB\,+\,BAO preference for low $\mnu$ in $\Lambda$CDM, or dynamical dark energy beyond it, both originate 
from the BAO preference for a low value of $\Omega_m$ \cite{Tang:2024lmo,Loverde:2024nfi}, the matter density in units of the critical density, which includes the neutrino contribution.  
With recent improvements in CMB lensing measurements, the CMB prefers a significantly higher value of $\Omega_m$ under the standard analysis, with the overall tension in $\Lambda$CDM rising to the 2.3$\sigma$ level if $\mnu=0.06$ \cite{DESI:2025ejh}.  
This can alternately be phrased as a preference for raising the CMB lensing amplitude through an artificial rescaling parameter $A_L$ \cite{Calabrese:2008rt} at a similar significance \cite{Green:2024xbb,SPT-3G:2024atg,Naredo-Tuero:2024sgf}.

Within the 6-parameter $\Lambda$CDM paradigm, the optical depth through reionization $\tau$ is the only parameter that could substantially change these conclusions and reconcile CMB, BAO, and neutrino oscillation measurements.  It has long been known that $\tau$ affects the interpretation of the lensing amplitude and hence the bound on $\mnu$ \cite{Smith:2006nk,Allison:2015qca}.  

The optical depth is also the most difficult of the $\Lambda$CDM parameters to measure, as it mainly involves the very lowest multipoles of the CMB E-mode polarization. It was first detected by WMAP in cross-correlation with temperature as
$\tau = 0.17 \pm 0.04$ \cite{WMAP:2003ggs}. Improvements in the measurements have steadily revised the bounds since its detection: from $\tau=0.089 \pm 0.014$ (WMAP9 \cite{WMAP:2012nax}) to
$\tau=0.097 \pm 0.038$ (Planck 2013  \cite{Planck:2013pxb}) to 
$\tau =0.079\pm 0.017$ 
(Planck 2015 \cite{Planck:2015fie})
to  $\tau=0.0544 \pm 0.00755$ (Planck 2018 \cite{Planck:2018vyg}).

These analyses assume a simple model of reionization where the ionization fraction changes nearly instantaneously from essentially neutral to full hydrogen ionization, such that reionization can be described by a single parameter. More generally, \lowLEE\ polarization data depends on 5 principal components of the ionization history
\cite{Hu:2003gh}. Taking this into account can both systematically raise the mean optical depth by allowing a high redshift component and broaden its modeling uncertainty. For example, with Planck 2015 data, the bound rises to $\tau=0.092\pm 0.015$ \cite{Heinrich:2016ojb}.
Furthermore, post Planck 2018, the SRoll2  reanalysis to improve foreground and systematic effects raises the bounds to 
$\tau= 0.0592 \pm 0.0062$ \cite{Delouis:2019bub} while still assuming a single-step reionization model. When combined with the principal-component analysis into RELIKE (Reionization Effective Likelihood), the value rises to $\tau=0.0619 \pm 0.0062$ \cite{Heinrich:2021ufa}. The revision to the Planck FlexKnot general reionization analysis \cite{Planck:2018vyg} (errata; v3) is also compatible with RELIKE. 
Finally, in models beyond slow-roll inflation, the \lowLEE\ polarization power can be suppressed to simultaneously accommodate the low large-angle temperature spectrum, requiring an even higher $\tau$ to match observations \cite{Obied:2018qdr}.

In light of these considerations, we conduct a thorough investigation of the role of $\tau$ and the leeway provided by data and modeling assumptions on  the tension between CMB\,+\,BAO \textit{within} $\Lambda$CDM given neutrino-oscillation measurements.  We begin in Sec.~\ref{sec:data} with the datasets and methodology. 
In Sec.~\ref{sec:tensions}, we systematically analyze the relationship between phrasing this tension in $\Lambda$CDM in terms of $\mnu$, $\Omega_m$, $A_L$, or $\tau$.  We address the leeway allowed by various analyses of $\tau$ in Sec.~\ref{sec:hightau}, showing that minimal-mass neutrinos can be reconciled with $\Lambda$CDM within the 95\% CL.  Dropping the \lowLEE\ constraint entirely completely removes the tension in $\Lambda$CDM and also weakens  the preference for dynamical dark energy with $\mnu=0.06$ to below 95\% CL. 
Finally, we discuss these results as well as the implications for replacing BAO with Type IA supernovae distance measures in Sec.~\ref{sec:discussion}.

Throughout this work, in addition to defining $M_\nu$ to omit the units of eV in the sum of the neutrino masses,
we also omit units for the current expansion rate $H_0\equiv 100 h$\,km/s/Mpc as a shorthand convention.

\section{Data and Methodology}
\label{sec:data}

We are primarily interested in reconciling CMB and BAO datasets with minimal neutrino masses within $\Lambda$CDM, but also test against the alternative of combining CMB and SN datasets.  
Our fiducial analyses therefore involve the following datasets: 
\begin{itemize}

    \item CMB: We use the Planck 2018 \cite{Planck:2018vyg} PR3 low-$\ell$ TT and EE and plik high-$\ell$ TTTEEE likelihoods.
    To these, we add Planck PR4 lensing \cite{Carron:2022eyg} and ACT DR6 lensing \cite{ACT:2023dou}, including their cross-correlations. 
    We also add SPT-3G MUSE lensing data \cite{SPT-3G:2024atg} as an independent data set since the cross-correlations are expected to be negligible considering the small but deep sky coverage \cite{ACT:2025rvn}. 
    
    \item BAO: We use BAO data from DESI DR2 \cite{DESI:2025zgx,DESI:2025ejh} with a likelihood based on Table 4 in \cite{DESI:2025zgx}, using $D_V/r_d$ for the lowest redshift bin and $D_M/r_d$, $D_H/r_d$, and their cross-correlation $r_{M,H}$ for all other bins. 
    
    \item SN: We use the DESY5 Supernovae distance moduli  \cite{DES:2024jxu}.
    
\end{itemize}

These datasets are the currently most constraining representations of their respective 
classes and cause the greatest tension in $\Lambda$CDM with their different $\Omega_m$ inferences \cite{Tang:2024lmo}.
In this work, we take them all at face value, but see Refs.~\cite{DESI:2025zgx,SPT-3G:2024atg,Naredo-Tuero:2024sgf,Reboucas:2024smm} for analyses of alternate subsets, likelihoods, and surveys.

Aside from $\mnu$, we restrict the parameter space to the six standard $\Lambda$CDM parameters: the physical cold dark matter density $\Omega_c h^2$; physical baryon density $\Omega_b h^2$; the angular scale $\theta_*$ of the sound horizon; the amplitude $A_s$ (varied as $\ln 10^{10} A_s$) and tilt $n_s$ of the initial curvature power spectrum; and the optical depth $\tau$ through reionization. We use broad, flat priors as well as all recommended nuisance parameters for the various datasets (e.g.\ \cite{Planck:2018vyg}). 
We use the approximation of 3 degenerate-mass neutrinos throughout, which for current cosmological data is sufficient to model the actual mass eigenstates of both the normal and inverted orderings 
\cite{Hannestad:2016fog,Herold:2024nvk}.

As we shall see, leeway in raising $\mnu$ to at least its minimum value is provided by relaxing the CMB lensing data, or almost equivalently, relaxing constraints on the optical depth through reionization $\tau$ from \lowLEE\ CMB polarization data. 
For lensing, we use the standard parameter characterizing the lensing tension, $A_L$, where the projected gravitational potential $C_L^{\phi\phi}$ is rescaled as $A_L C_L^{\phi\phi}$ in parameter space for all CMB power spectra; separately we consider the results from omitting the CMB lensing reconstruction data. 

We are particularly interested in exploring the robustness of the tension to constraints on $\tau$ from different data, analysis, and modeling choices. We first drop all \lowLEE\ data and constraints by removing it from the CMB likelihoods.  We then test against the SRoll2 modifications to the low-$\ell$ EE Planck 2018 data and analysis that improve the foreground and systematic errors \cite{Delouis:2019bub}.  

Finally, we test the standard modeling of reionization and inflation. For reionization, we take the $\tau$ constraint from the principle-component analysis of RELIKE using the SRoll2 data.  For inflation, we follow Planck 2013-2018 in testing an exponential suppression of large-scale  curvature power ${\cal P}_{\cal R}$ 
by multiplying the initial curvature spectrum  by  \cite{Planck:2013jfk} 
\begin{equation}
    1-e^{-\sqrt{k/k_c}}\,,
    \label{eq:kc}
\end{equation}
with a flat prior on
$\ln (k_c\,{\rm Mpc}) \in [-12,-3]$.

These modifications from the baseline CMB analysis are denoted as
CMB\,(X), where the modification X is given by:
\begin{itemize}

    \item CMB\,($A_L$): lensing power amplitude $A_L$ marginalized over for both lensing reconstruction and the smearing of peaks in CMB primaries 
    
    \item CMB\,(no lens recon): no CMB lensing reconstruction from Planck PR4, ACT and SPT
    
    \item CMB\,(no \lowLEE): no \lowLEE\ constraint of any kind
    
    \item CMB\,(\lowLEE): $\tau=0.0544 \pm 0.00755$ prior replaces \lowLEE\ data \cite{Planck:2018vyg,Heinrich:2021ufa}
    
    \item CMB\,(SRoll2): $\tau= 0.0592 \pm 0.0062$ prior \cite{Delouis:2019bub}
    
    \item CMB\,(RELIKE): $\tau=0.0619 \pm 0.0062$ prior \cite{Heinrich:2021ufa}
    
    \item CMB\,(${\cal P}_{\cal R}$): curvature power suppression scale $k_c$ marginalized with SRoll2 data and standard reionization

\end{itemize}

Due in part to the preference for less than minimal mass neutrinos with CMB\,+\,BAO in $\Lambda$CDM, we use a mix of Bayesian and frequentist methods in our analysis.  
For both methods, we use the Boltzmann code \texttt{CLASS} (Cosmic Linear Anisotropy Solving System) to compute the power spectra and distances \cite{Blas:2011rf}. 
We use functionality already built into \texttt{CLASS} to provide an externally calculated array for the initial curvature power spectrum for power-suppression tests. For our Bayesian explorations, we use Metropolis-Hastings Markov Chain Monte Carlos (MCMC) with \texttt{MontePython} \cite{Audren:2012wb,Brinckmann:2018cvx} and \texttt{Cobaya} \cite{Torrado:2020dgo}. We use the Python package \texttt{GetDist} \cite{Lewis:2019xzd} to analyze the chains.

While Bayesian methods are more efficient at providing multi-dimensional parameter constraints and incorporating prior knowledge,
to address the preference in the data for less than minimal mass neutrinos, we 
use \texttt{Procoli} \cite{Karwal:2024qpt} to obtain likelihood profiles and optimize $\chi^2$ values to compare model parameters. 
\texttt{Procoli} uses simulated-annealing optimizers and has been employed in other instances where prior effects are important \cite{Naredo-Tuero:2024sgf,Poulin:2023lkg, Poudou:2025qcx,Chatrchyan:2024xjj,Greene:2024qis,Efstathiou:2023fbn}.

We report $\Delta\chi^2$ values for various datasets against a common model where $\mnu=0.06$ to illustrate the absolute improvements in the fit.  When changing the assumptions involving lensing and $\tau$, we isolate the portion that comes from data excluding all \lowLEE\ information (X=no \lowLEE) and denote this as
$\Delta\chisqnoEE$.

\section{Shifting tensions: $\mnu$, $\Omega_m$, $A_L$, $\tau$}
\label{sec:tensions}

We show here that the tension between CMB and BAO datasets, which originates from the BAO preference for low $\Omega_m$ in $\Lambda$CDM, can be alternately phrased as tensions with minimal neutrino masses (Sec.~\ref{sec:mnu}), CMB lensing amplitude (Sec.~\ref{sec:AL}) or the optical depth through reionization (Sec.~\ref{sec:tau}).  
While lowering the neutrino mass or raising the lensing amplitude is not physically possible in $\Lambda$CDM, the optical depth is a physical parameter that can be raised were it not for the CMB \lowLEE\ constraints in the standard analysis.

\subsection{Neutrino masses $\mnu$}
\label{sec:mnu}

We begin with the usual way of casting the tension between CMB and BAO datasets in $\Lambda$CDM as a preference for neutrino masses that are significantly below the minimum allowed by oscillation experiments ($\mnu=0.06$) and even formally negative \cite{Craig:2024tky,Green:2024xbb,Loverde:2024nfi,DESI:2025ejh,Lynch:2025ine,Naredo-Tuero:2024sgf,RoyChoudhury:2024wri, Reboucas:2024smm,Jiang:2024viw,Herold:2024nvk}.
\begin{figure}
    \centering
    \includegraphics[width=0.95\linewidth]{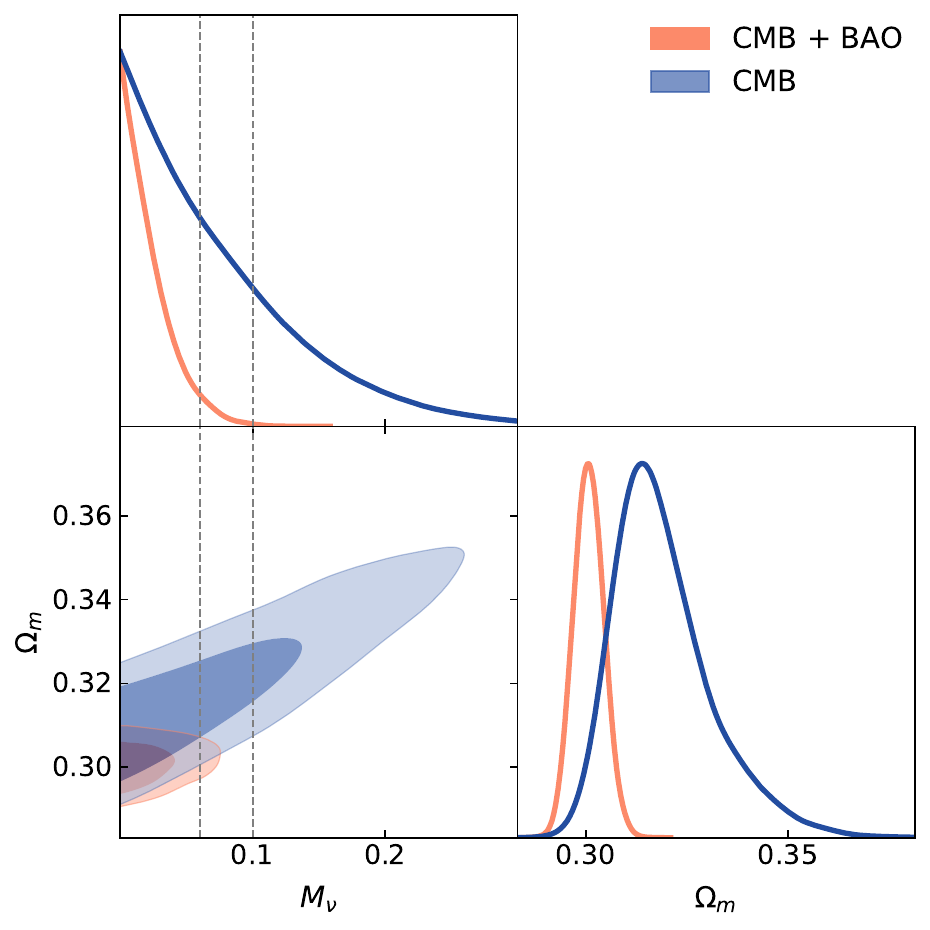}
    \vskip -0.5cm
    \caption{$\mnu-\Omega_m$ posterior constraints for CMB and CMB\,+\,BAO datasets.  Along the CMB degeneracy of $\Delta M_\nu \approx 2.5 \Delta \Omega_m/\Omega_m$\,, the BAO data favor $\mnu$ to be below the minimal values of $0.06$ and $0.1$ for normal and inverted ordering respectively (vertical dashed lines).}
    \label{fig:mnuOmegam}
\end{figure}

This tension originates from the BAO preference for a low value of $\Omega_m = 0.2975 \pm 0.0086$ \cite{DESI:2025zgx}  
when combined with the well-measured angular scale $\theta_*= r_s/D_A(z_*)$ of the CMB acoustic peaks, where $r_s$ is the sound horizon and $D_A$ is the angular diameter distance to the redshift of recombination $z_*$. Neutrinos that are relativistic at recombination count as radiation for $r_s$ and matter for $D_A$ and $\Omega_m$, which dictates their role in geometric constraints (e.g.\ \cite{Loverde:2024nfi}). 
To the extent that the energy densities of various components at recombination are fixed, $r_s$ is fixed in Mpc$^{-1}$. $D_A(z_*)$ is then also fixed by the precise measurement of $\theta_*$. Thus, the addition of the neutrino matter density at low redshift $\rho_\nu \propto (1+z)^3\Omega_\nu h^2 \propto (1+z)^3 \mnu/93.1$\, implies that the Hubble constant $H_0$ must decrease as $\mnu$ increases to compensate the change in distance (e.g.~\cite{Planck:2013pxb}).  This leads to the strong scaling between $\mnu$ and $\Omega_\nu$\,, and consequently $\Omega_m$, at fixed CMB geometry despite the small fraction of the matter contained in neutrinos for sub-eV masses (e.g.~\cite{Loverde:2024nfi}).

This can be seen in Fig \ref{fig:mnuOmegam}, where we show posteriors for $\mnu$ and $\Omega_m$ with and without BAO data. CMB data alone follow an approximate degeneracy which allow changes of $\Delta \mnu \approx 2.5\Delta\Omega_m/\Omega_m$. This degeneracy is restricted by the BAO preference for low $\Omega_m$. To compensate for this, $\mnu$ decreases below its minimum possible value. Without the BAO data and a physical prior of $\mnu\ge 0$, 50\% of the $\mnu$ posterior lies above $\mnu=0.06$\,, whereas with BAO only 5\% does.  The inverted-ordering minimal mass of $0.1$eV\, would then be strongly ruled out. Moreover, with the BAO data, the posterior peaks sharply at $\mnu=0$\,, implying that the preferred neutrino mass would be negative if that were physically possible.

\begin{figure}
    \centering
    \includegraphics[width=0.97\linewidth]{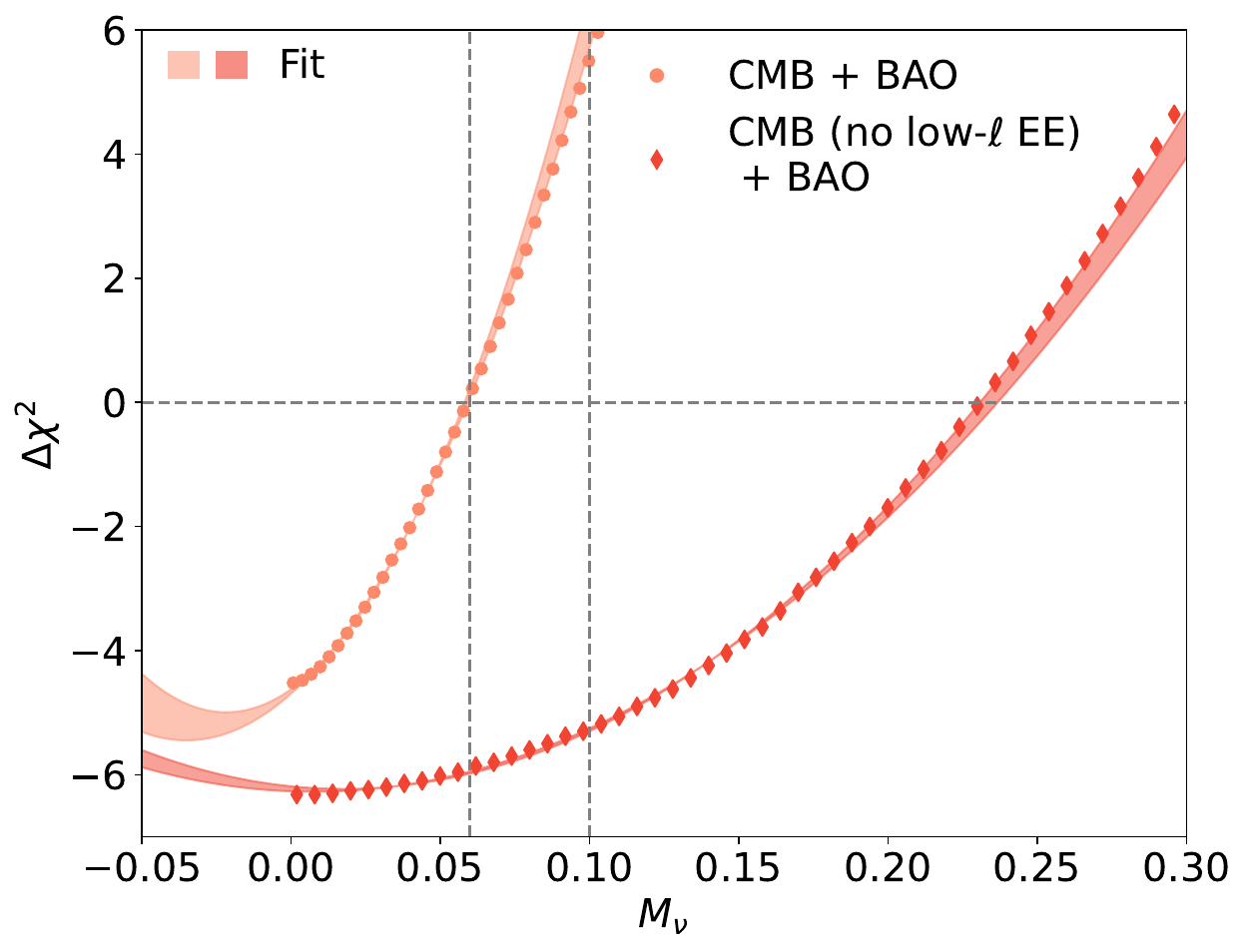}
      \vskip -0.25cm
    \caption{Likelihood profiles shown as $\Delta\chi^2$ along $\mnu$ for CMB\,+\,BAO and CMB\,(no \lowLEE)+BAO datasets with the zero point defined by the CMB\,+\,BAO $\mnu=0.06$\, model in Tab.~\ref{tab:chi2} (horizontal dashed line). 
    While CMB\,+\,BAO are in tension with both the normal and inverted-ordering minimal masses (vertical dashed lines), removing the optical depth $\tau$ information from \lowLEE\ restores consistency with both.  Shaded bands represent a range of quadratic fits to extrapolate the profile to $\mnu<0$ (see text).
    }
\label{fig:profile}
\end{figure}

In Fig \ref{fig:profile}, we ascertain this by extrapolating a profile likelihood for $M_\nu$ to the unphysical negative mass regime with CMB\,+\,BAO. To extrapolate the profile below $\mnu = 0$ and find the minimum, we perform a quadratic fit. We find that the minimum lies between $-0.035 \leq M_\nu \leq -0.022$.  The  upper and lower values come from fitting the profile points with $\Delta \chi^2 \leq 4$ and $\leq 9$ above the physical minimum respectively and reflects the uncertainty in our extrapolation.
For comparison, using a parameterized extension to $\mnu{}_{\rm ,eff}<0$, Ref.~\cite{DESI:2025ejh} obtains 
$\mnu{}_{\rm ,eff}=-0.101^{+0.047}_{-0.056}$
for a case where their profile extrapolation gives $\mnu = -0.036$. 

\renewcommand{\arraystretch}{1.2} 
\begin{table*}[ht]
\centering
\begin{tabularx}{0.62\textwidth} { 
  | >{\centering\arraybackslash}X 
   >{\centering\arraybackslash}X 
  | >{\centering\arraybackslash}X 
  | >{\centering\arraybackslash}X 
   >{\centering\arraybackslash}X 
   >{\centering\arraybackslash}X | }
 \hline
  & $\mnu$ & $\tau$ & $\Delta\chi^2$ & $\Delta\chi^2_\tau$ & $\Delta\chisqnoEE$    \\
 \hline
CMB\,+\,BAO  & 0.06  & 0.065   & 0  & 0 &  0\\
&  0 & 0.061 & -4.52 &-1.65  &-2.87  \\
\hline
$A_L$  & 0.06  & 0.050  & -10.05  & -3.64 & -6.41 \\
& 0 & 0.050  &  -10.42  & -3.63 & -6.79  \\
\hline
no \lowLEE &  0.06  &  0.092   &  -5.87 & -- & -5.87 \\
& 0 & 0.082  & -6.32 & --  & -6.32\\
\hline
\lowLEE &  0.06  & 0.066   & 0.10  & 0.33  & -0.23 \\
& 0 & 0.063 & -4.22 &-0.72  & -3.49\\
\hline
SRoll2 &  0.06  & 0.066  & 0.05  & 0.30 & -0.26 \\
& 0 & 0.064  &-4.25 & -0.25 & -4.00 \\
\hline
RELIKE &  0.06 &0.070    & -0.45  & 1.23 & -1.68 \\
& 0 &0.066   & -4.30 & 0.21  &-4.51\\
\hline
${\cal P}_{\cal R}$  &  0.06 & 0.070     & -1.96  & 0.99 & -2.96 \\
& 0 &0.064    & -6.41 & -0.91  & -5.50\\
\hline
\end{tabularx}
\caption{$\Delta \chi^2$ for the relaxation of tension between minimal neutrino masses $\mnu=0.06$ and $0$.  All $\Delta\chi^2$ values are relative to the CMB\,+\,BAO $\mnu=0.06$ model and minimized with respect to $\Lambda$CDM parameters for each of X 
alterations in datasets and methodology: CMB\,(X)+BAO with X=$A_L$ for optimizing the lensing-rescaling parameter; no \lowLEE\ for dropping all \lowLEE\ constraints; \lowLEE\ for replacing the data with the corresponding $\tau$ prior; SRoll2 for its  $\tau$ prior; RELIKE for SRoll2 priors with general reionization models; ${\cal P}_{\cal R}$ for an exponential truncation of the power spectrum at $k_c$.  
$A_L$ and no \lowLEE\ cases resolve the tension similarly with
optimized values of $A_L=1.09$, $1.07$ or $\tau=0.092,0.082$ for $\mnu=0.06$, 0 respectively. 
For, ${\cal P}_{\cal R}$ the optimized values are $\ln (k_c \, {\rm Mpc})= -8.56 \,, -8.75$ respectively and we use the SRoll2 likelihood for $\Delta\chi^2_\tau$. See Sec.~\ref{sec:data} for details.
}
\label{tab:chi2}
\end{table*}

In Tab.~\ref{tab:chi2}, we also extract the $\Delta\chi^2$ between the $\mnu=0.06$\, and the physical minimum $\mnu=0$\, cases, where all other parameters have been adjusted to minimize $\Delta\chi^2$. Note that the values of $\Delta \chi^2$ for all cases in Tab.~\ref{tab:chi2} are given with respect to the CMB\,+\,BAO fiducial model with $\mnu=0.06$ in order to compare the goodness of fit. 
For CMB\,+\,BAO 
\begin{equation}
    \chi^2(\mnu=0) -\chi^2(0.06)\approx -4.5, 
    \label{eq:CMBBAOchi2}
\end{equation}
and a substantial part of this preference for lowering $\mnu$ comes from the contribution of the \lowLEE\ likelihood $\Delta\chi_\tau^2\approx -1.7$.  
This already highlights the role of the optical depth $\tau$ in this tension as we shall explore in detail below.

\begin{figure}
    \centering
    \includegraphics[width = 0.9\linewidth]{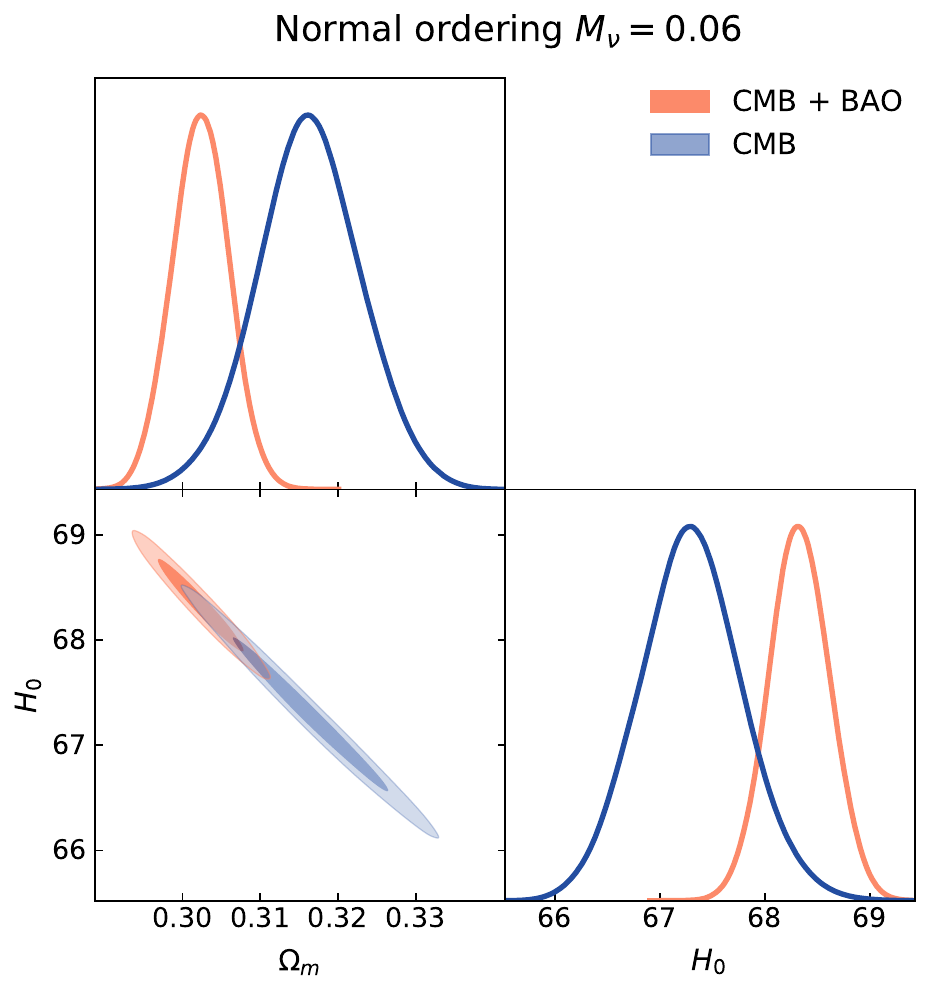}
    \caption{$\Lambda$CDM geometric tension in $\Omega_m-H_0$ at the minimal $\mnu=0.06$.
    With neutrino masses fixed, the CMB geometric degeneracy is $\Delta\Omega_m/\Omega_m \approx -3 \Delta H_0/H_0$ and tension between $\Omega_m$ values with CMB and CMB\,+\,BAO reflect the calibration of the sound horizon in the former. 
    }
\label{fig:OmegamH0}
\end{figure}

Since we cannot reconcile CMB and BAO data within $\Lambda$CDM with $\mnu$ constraints from oscillation experiments, another way to do so is to allow more freedom in the calibration of the sound horizon $r_s$. In Fig.~\ref{fig:OmegamH0}, we fix $\mnu=0.06$ and highlight the problem in the $\Omega_m-H_0$ plane. 
Throughout the figures here and below, when we have fixed $\mnu$ to the minimal value required in normal ordering and shifted the tension elsewhere, we use the label ``normal ordering $\mnu=0.06$."
With neutrinos fixed, the CMB geometric constraint on $\theta_*$ requires constant $\Omega_m h^3$\,, and this degeneracy is broken by measurements of the cold dark matter density $\Omega_c h^2$ and baryon density $\Omega_b h^2$ that calibrate $r_s$ \cite{Hu:2000ti}. Adding the BAO constraint pulls $\Omega_m$ down and $H_0$ up, reflecting the same tension that drives the neutrino mass constraint but now directly in $\Omega_m$. With CMB\,+\,BAO at $\mnu = 0.06$, $\Omega_m = 0.3024\pm 0.0036$,  whereas with only CMB $\Omega_m = 0.3163\pm 0.0067$  which is a $2.5\sigma$ 
tension in the update difference in means (Eq. (53) in \cite{Raveri:2018wln}). 
We shall see that the information from CMB lensing and consequently $\tau$ plays a substantial role in determining this calibration.

\subsection{CMB lensing rescaling $A_L$}
\label{sec:AL}

CMB lensing data plays a significant role in the tension between CMB and BAO datasets \cite{Green:2024xbb}, and its significance has increased with the recent ACT and SPT measurements \cite{ACT:2023dou,SPT-3G:2024atg}. This tension can be quantified as a preference for excess lensing $A_L>1$, where $A_L$ is a parameter that rescales the lensing power spectrum as $C_L^{\phi\phi}\rightarrow A_L C_L^{\phi\phi}$ for each set of $\Lambda$CDM parameters \cite{Calabrese:2008rt}.
Note that when comparing $A_L$ values with $\Omega_m$ marginalized or maximized over across different datasets, the interpretation of $A_L$ for the lensing power spectrum itself changes, but the results can still be compared to the expectation that $A_L=1$.

The relationship between the geometric CMB\,+\,BAO constraints and lensing comes about from the need to compensate the changes due to low $\Omega_m$. Compensation  in  $C_L^{\phi\phi}$ occurs along the geometric $\mnu-\Omega_m$ degeneracy, since lowering $\mnu$ removes the suppression of $C_L^{\phi\phi}$ below the neutrino free-streaming scale whereas lowering $\Omega_m$ suppresses the gravitational potential fluctuations responsible for lensing.  

At fixed $\theta_*$, the relationship between a change in $\mnu$ and the change in lensing is roughly $\Delta \mnu \approx 2.5 \Delta \ln C_L^{\phi\phi}$ (\cite{Smith:2006nk}, their Fig. 4 below the free-streaming scale).
This implies that to restore the minimal mass of $0.06$ from 0, we would need to raise the amplitude of lensing by a factor of $\sim 1.024$, which agrees with the shift in $A_L$ in Tab.~\ref{tab:chi2}. Extrapolating the degeneracy line to bring the preferred value back to $A_L=1$ would require $\mnu\approx -0.15$. This is another way of extrapolating to negative neutrino mass and it provides a more negative value than the profile extrapolation in Fig.~\ref{fig:profile}.

\begin{figure}
     \centering
\includegraphics[width=0.45\textwidth]{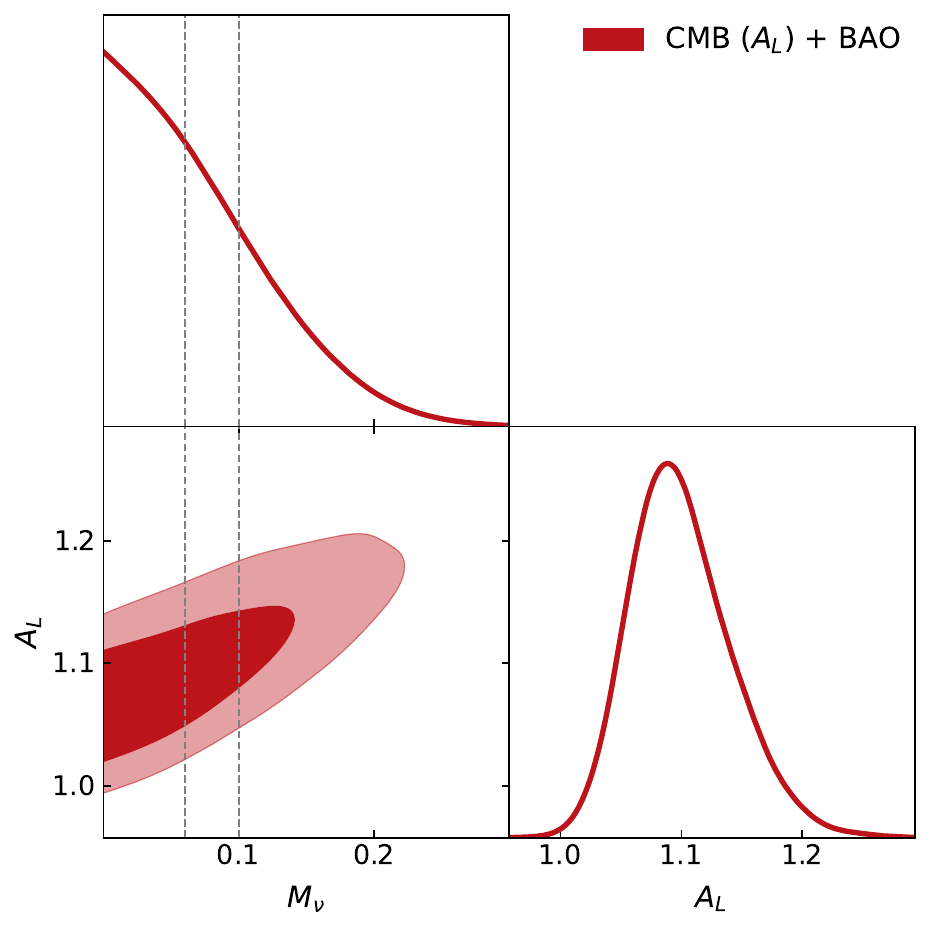}
 \vskip 0.01cm
\hskip -0.16cm \includegraphics[width=0.46\textwidth]
 {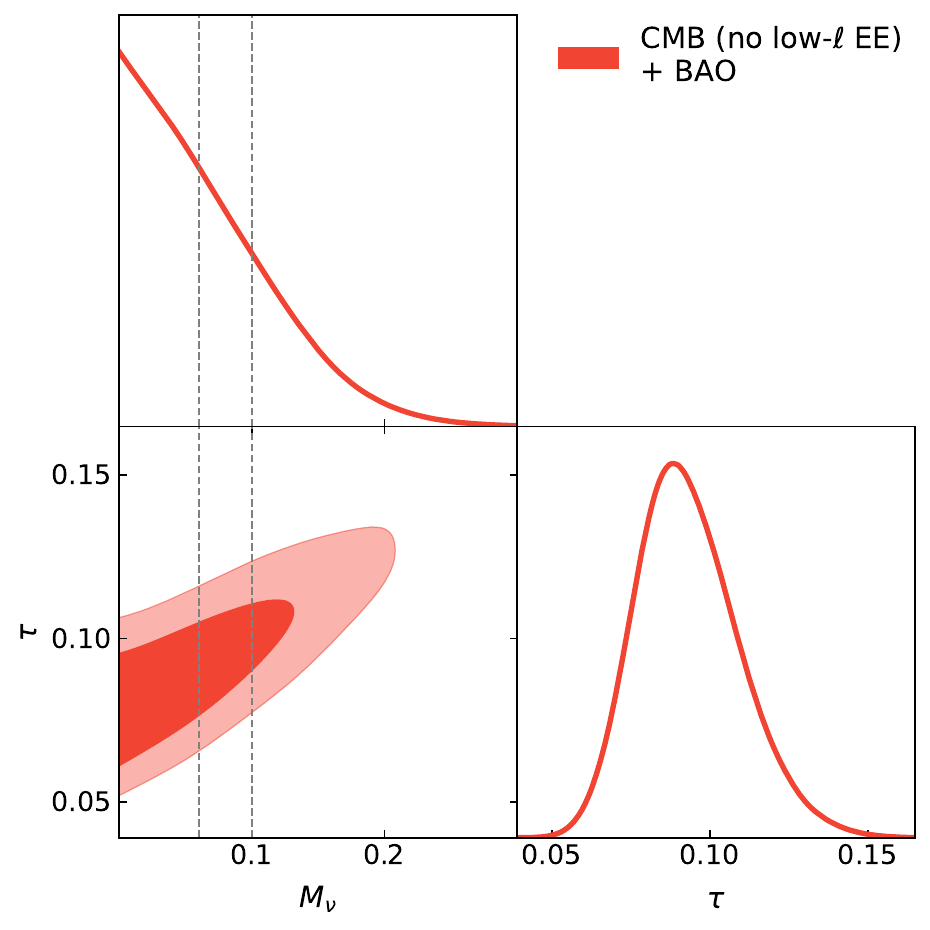}
  \vskip -0.1cm
     \caption{CMB\,+\,BAO tension with minimal neutrino mass from normal and inverted orderings (vertical lines) is relaxed by either rescaling lensing with $A_L$ or by removing the \lowLEE\ constraint on $\tau$ due to the degeneracy
     $\Delta M_\nu  \approx 5\Delta \tau \approx 2.5\Delta \ln A_L$.}
    \label{fig:mnuALtau}
\end{figure}

\begin{figure}
    \centering
    \includegraphics[width = 0.95\linewidth]
    {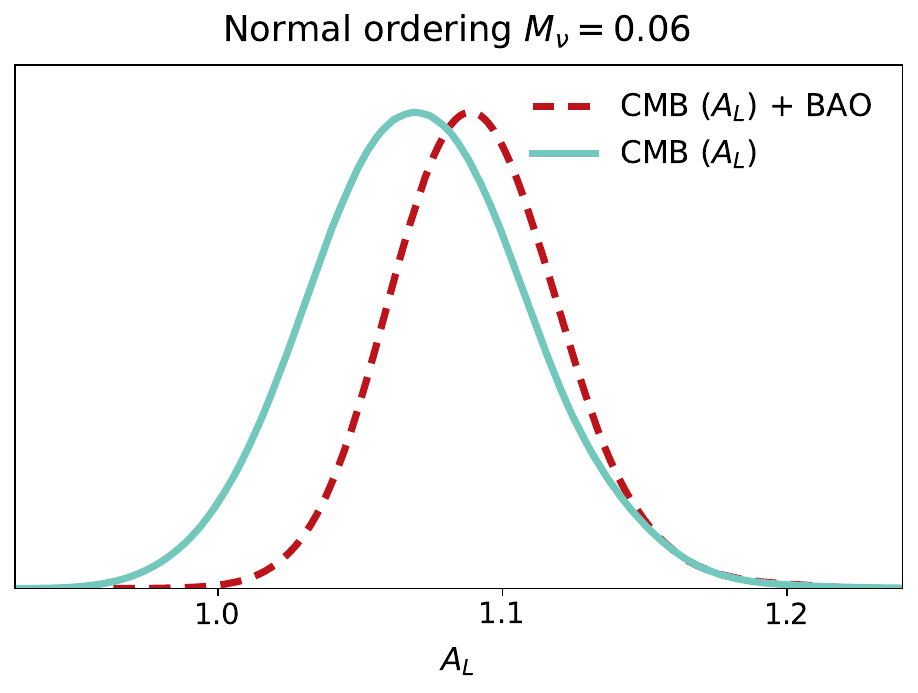}
    \vskip -0.2cm
    \caption{Posterior constraints on lensing-rescaling parameter $A_L$ with CMB and CMB\,+\,BAO.  With the addition of BAO, $A_L$ becomes strongly inconsistent with the $\Lambda$CDM value of $A_L=1$, with a maximum likelihood value of $A_L=1.09$.
    }
\label{fig:AL}
\end{figure}
  
We see this degeneracy between $\mnu$ and $A_L$ in the top panel of Fig \ref{fig:mnuALtau}. 
While in the $\mnu-A_L$ plane at $\mnu \rightarrow 0$, $A_L=1$ is still barely allowed at $95\%$ CL, for $\mnu=0.06$ the exclusion is much higher. 
In Fig.~\ref{fig:AL} for $\mnu=0.06$, we show that  with CMB data alone 3\% of the posterior lies below $A_L=1$ whereas with CMB+BAO this value is too far in the tails to reliably measure with the MCMC.  
Moreover, it has been shown that this tension is now driven by $C_L^{\phi\phi}$ measurements rather than the smoothing of the acoustic peaks \cite{Green:2024xbb,SPT-3G:2024atg}.

\begin{figure}
    \centering
    \includegraphics[width = 0.95\linewidth]{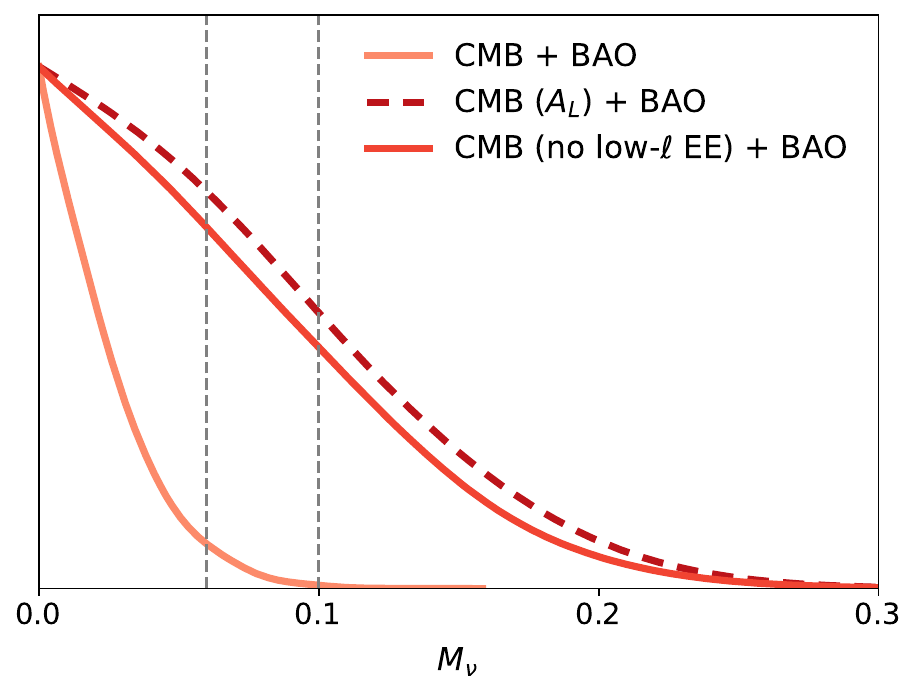}
    \vskip -0.3cm
    \caption{Posterior distributions for $\mnu$ for CMB\,+\,BAO vs.~CMB\,($A_L$)+BAO and CMB\,(no \lowLEE)+BAO.   
    The $A_L$ and no \lowLEE\ extensions resolve the tension with minimal masses (vertical lines) similarly.
}
\label{fig:mnu_tension_resolution}
\end{figure}

When marginalizing $A_L$ in Fig.~\ref{fig:mnu_tension_resolution}, $\mnu\ge 0.06$ represents 52\% of the posterior compared to just $5\%$  without $A_L$. Correspondingly  in Tab.~\ref{tab:chi2}, for
CMB\,($A_L$)\,+\,BAO, 
\begin{equation}
    \Delta\chi^2(\mnu=0) -\Delta\chi^2(0.06)\approx -0.4 \,,
\end{equation}
which is a negligible difference compared with that of CMB\,+\,BAO in Eq.~(\ref{eq:CMBBAOchi2}). 
The overall improvement in $\Delta\chi^2\sim -10$ reflects a large component from removing the optical depth penalty of $\Delta\chi^2_\tau\approx 3.6$ in the fiducial model. 
The values of $\Delta\chisqnoEE \sim -6.4$ and $\sim -6.8$ for $\mnu=0.06$ and $\mnu=0$ respectively
reflect the part of this improvement from omitting the \lowLEE\ data. As we shall see below, this represents the maximum improvement possible by relaxing $\tau$ constraints.  Notice that this is achieved by setting $\tau\approx 0.05$ which is slightly below the 0.0544 central value of the CMB\,(\lowLEE) (see Sec.~\ref{sec:data}) constraint.   This is because  part of this constraint comes from the peak smoothing anomaly which is removed by marginalizing $A_L$.  The Planck 2018 primary CMB constraint with $A_L$ marginalized is 
$\tau = 0.0492^{+0.0088}_{-0.0073}$ \cite{Planck:2018vyg}.  

Of course, $A_L$ is an artificial  parameter designed to monitor tension between the lensed and unlensed CMB in $\Lambda$CDM. In the context of $\Lambda$CDM, the strong constraint on the amplitude of $C_L^{\phi\phi}$ with CMB\,+\,BAO instead provides a constraint on $\Omega_c h^2$ given the high sensitivity at fixed $\theta_*$:
$\Delta \ln C_L^{\phi\phi} \sim 20 \Delta \Omega_c h^2$ \cite{Smith:2006nk}.  CMB lensing now forms a substantial part of the constraint on the sound horizon $r_s$ involved in the CMB\,+\,BAO geometric tension with $\Omega_m$. Relaxing this by allowing the lensing amplitude to vary then shifts the neutrino tension to a lensing tension.

\subsection{Optical depth $\tau$}
\label{sec:tau}

While adjusting the lensing amplitude to accommodate the lensing tension through $A_L$ is unphysical, there is a standard $\Lambda$CDM parameter that has the same effect: the optical depth $\tau$ through reionization.  
Measurements of the acoustic peak amplitudes constrain $A_s e^{-2\tau}$, where $A_s$ is the amplitude of the initial curvature power spectrum.  Raising $\tau$ increases the inferred $A_s$ and hence the amount of lensing. In fact, measurement and modeling uncertainties on $\tau$ have long been known to be an important issue for measuring $\mnu$ (e.g., \cite{Smith:2006nk,Allison:2015qca}).  From the perspective of lensing measurements, this amplitude degeneracy can be cast as a degeneracy with $A_L$ that leaves $A_L e^{-2\tau}$ fixed.

In the absence of \lowLEE\ data, this resolution would work equally well as $A_L$.  In Fig.~\ref{fig:mnuALtau} (lower panel) we show the
posterior constraints in $\mnu-\tau$ for 
CMB\,(no \lowLEE)\,+\,BAO.   Notice that the degeneracy line follows
\begin{equation}
\Delta \mnu \approx 5 \Delta\tau  \approx 2.5 \Delta \ln A_L
\end{equation}
as expected and so a shift from $\tau\sim 0.054$ to $0.092$ works comparably well as $A_L\sim 1.09$ in  Tab.~\ref{tab:chi2}.  More specifically,
with CMB\,(no \lowLEE)+\,BAO
\begin{equation}
    \Delta\chi^2(\mnu=0) -\Delta\chi^2(0.06)\approx -0.5,
\end{equation}
and is again insignificant when compared with that of CMB\,+\,BAO in Eq.~(\ref{eq:CMBBAOchi2}).  Moreover, for both $\mnu$ values, the $\Delta\chisqnoEE \sim -6$  is comparable to that with $A_L$ optimized, reflecting a similar level of relaxation of the CMB\,+\,BAO tension but now within $\Lambda$CDM.  
Correspondingly, $\mnu\ge 0.06$ is allowed and is $48$\% of the posterior in Fig.~\ref{fig:mnu_tension_resolution} (bottom panel). 
Furthermore, $\mnu \ge 0.1$ is $25\%$ of the posterior such that the inverted ordering is no longer ruled out.
In Fig.~\ref{fig:profile}, the $\mnu$ profile peaks near $\mnu=0$ and $\mnu = 0.06$ remains in the flat portion of the curve.

\begin{figure}
    \centering
    \includegraphics[width = 0.95\linewidth]{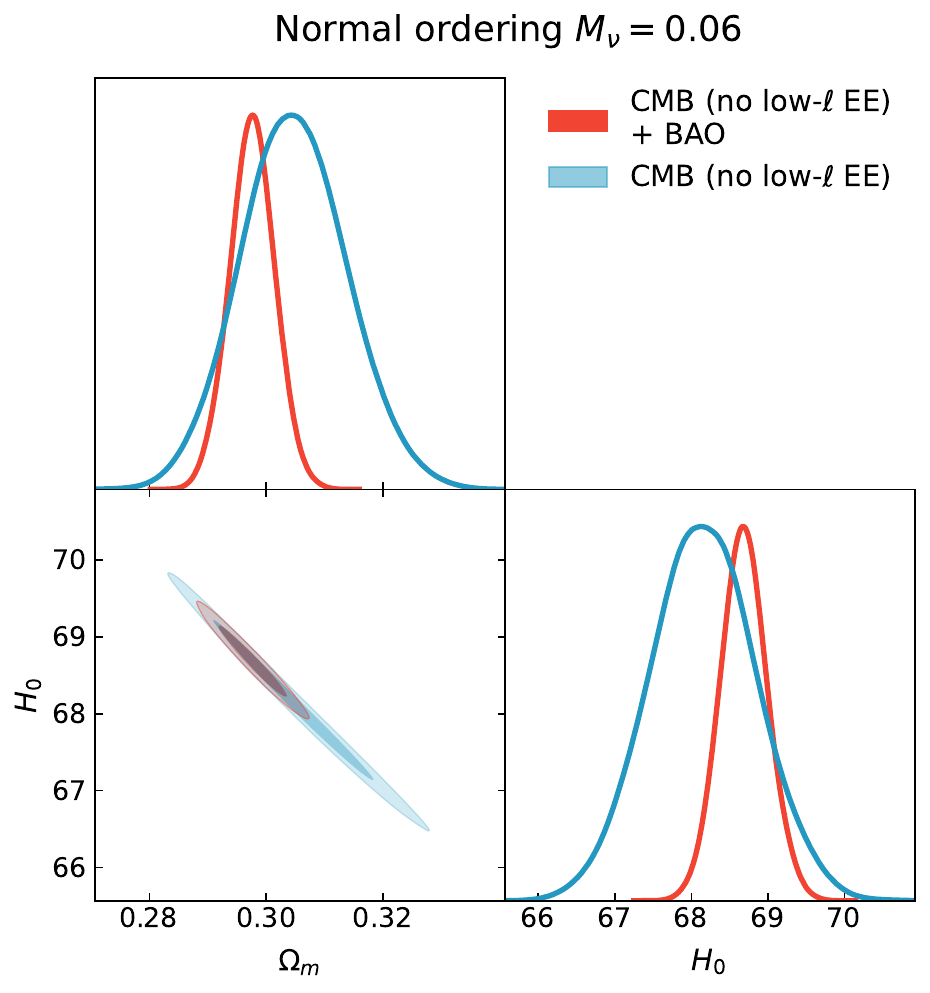}
    \vskip -0.2cm
    \caption{Resolving the $\Lambda$CDM geometric tension in $\Omega_m-H_0$ at fixed $\mnu=0.06$ with $\tau$ (cf.~Fig.\,\ref{fig:OmegamH0}).  With CMB\,(no \lowLEE), constraints on the sound horizon $r_s$ weaken and allow for the low $\Omega_m$ and high $H_0$ preferred by BAO.
    }
\label{fig:OmegamH0nolowLEE}
\end{figure}

Finally with fixed $\mnu=0.06$, we obtain the following posterior constraints: 
\begin{eqnarray}
    \tau &=& 0.091 \pm 0.011 \quad \mbox{CMB\,(no \lowLEE)+BAO}, \nonumber\\
    &=& 0.080 \pm 0.016 \quad \mbox{CMB\,(no \lowLEE).} 
\end{eqnarray}
We can trace the physical reason for the change in CMB geometric inferences back to the $\Omega_m-H_0$ plane in Fig.~\ref{fig:OmegamH0nolowLEE}.  Compared with Fig.~\ref{fig:OmegamH0}, the impact of omitting \lowLEE\ information is a substantial broadening of the allowed range of $\Omega_m$, or equivalently, the calibration of $r_s$ that now allows the favored low $\Omega_m$ without lowering $\mnu$. Along the $\theta_*$ degeneracy, the additional consequence is that larger $H_0$ is allowed by the CMB and preferred by BAO
\begin{eqnarray} 
H_0&=& 68.69\pm 0.31 \quad \mbox{CMB\,(no low-$\ell$ EE)+BAO}, \nonumber\\
&=& 68.15\pm 0.69 \quad \mbox{CMB\,(no low-$\ell$ EE)}.
\end{eqnarray}

\begin{figure}
    \centering    
    \includegraphics[width=0.97\linewidth]{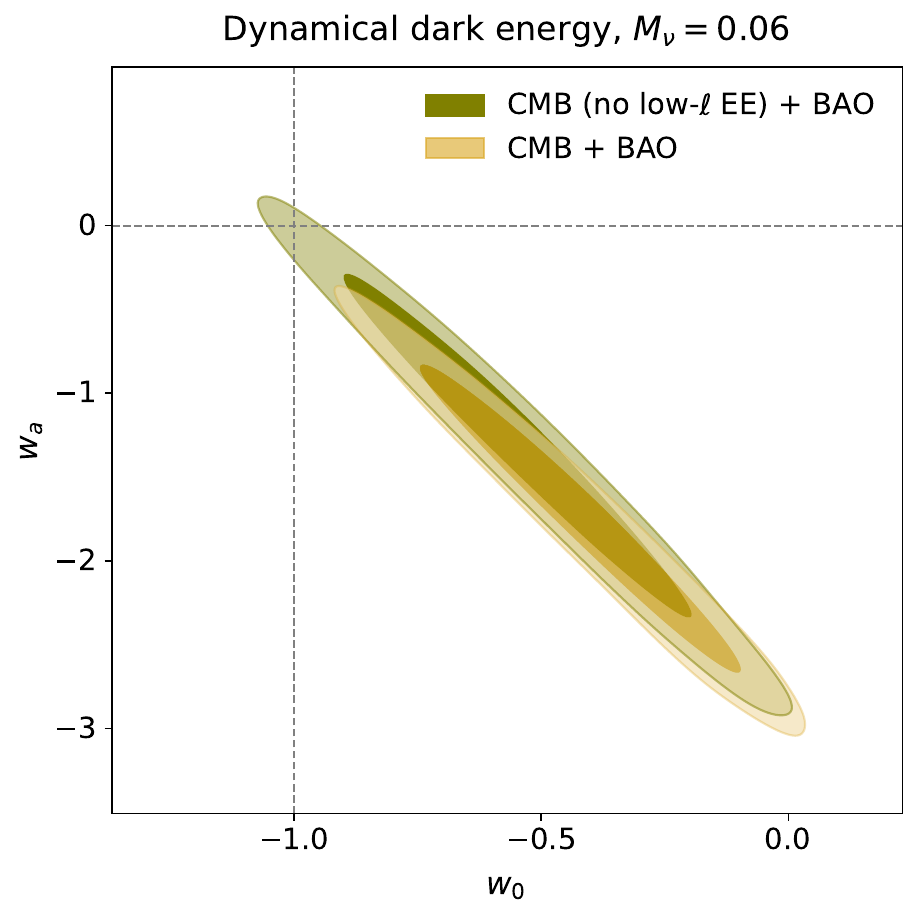}
    \caption{Preference for dynamical dark energy  with $\mnu=0.06$ drops from substantially above $95\%$ CL in CMB+BAO  to below $95\%$ CL in CMB\,(no \lowLEE)\,+\,BAO. Note that $\Lambda$CDM lies at the intersection between the two dashed lines at $w_0=-1,w_a=0$. 
    }
    \label{fig:w0wa}
\end{figure}

Given that CMB\,(no \lowLEE)+BAO is compatible with $\Lambda$CDM and minimal $\mnu$, removing the \lowLEE\ data also reduces the preference for dynamical dark energy in these datasets.   In Fig.~\ref{fig:w0wa}, we show the usual $w_0-w_a$ dynamical dark energy extension where the dark energy equation of state is $w(a)= w_0+w_a(1-a)$.   With CMB\,+\,BAO, the $\Lambda$CDM model (where $w_0=-1,w_a=0$) is excluded at greater than $95\%$ CL whereas with
CMB\,(no \lowLEE) it is within the $95\%$ confidence region.
Conversely, it is known that beyond $\Lambda$CDM, due to the degeneracy in $C_L^{\phi\phi}$ with dark energy \cite{Smith:2006nk}, the neutrino mass tension relaxes and is insignificant if $w_0-w_a$ is marginalized \cite{DESI:2025ejh}. 

\begin{figure}
    \centering
    \includegraphics[width = 0.97\linewidth]{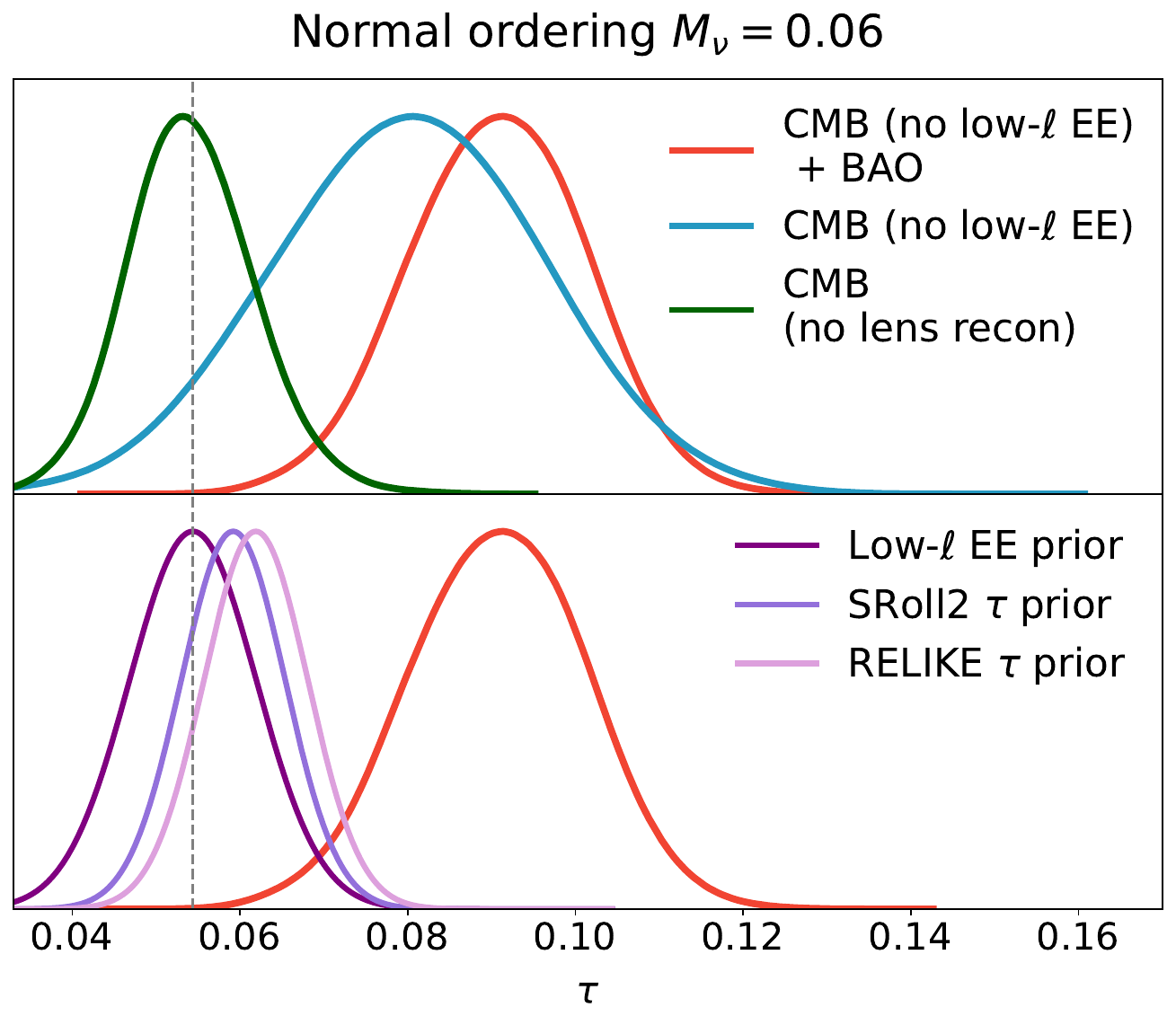}
    \vskip -0.25cm
    \caption{Prior and posterior distributions for $\tau$ for the minimal $\mnu=0.06$.
    Top panel: CMB\,(no \lowLEE) and CMB\,(no \lowLEE)+BAO are compatible at higher values of $\tau$ but in tension with the \lowLEE\ constraint from the primary CMB anisotropies (no lens reconstruction).  
    Bottom panel: priors on $\tau$ from alternate analyses: \lowLEE\ prior that 
    matches its constraint (vertical line highlights the alignment given the  small asymmetry of the data constraint); RELIKE extended reionization models with SRoll2.  RELIKE $\tau$ prior partially relaxes the tension with BAO.
    }
    \label{fig:tau}
\end{figure}

Of course, the actual \lowLEE\ data disfavor this $\tau$ resolution in $\Lambda$CDM. 
In Fig.~\ref{fig:tau} (top panel) we compare the \lowLEE\ constraint on $\tau$ from the posterior of the CMB analysis without lensing reconstruction data to these inferences.   
The $1.4\sigma$ level agreement between CMB\,(no \lowLEE) and the \lowLEE\ data becomes a $2.7\sigma$ tension with the inclusion of BAO.  This tension motivates a study of the robustness of the \lowLEE\ constraint on $\tau$ next.

\section{Elevating the optical depth} 
\label{sec:hightau}

Within the $\Lambda$CDM model, the way of reconciling CMB and BAO data without violating neutrino-oscillation constraints is to raise the optical depth $\tau$.   
On the other hand, as we have seen in the previous section, the tension with the \lowLEE\ measurement of $\tau$ is at $2.7\sigma$ compared with CMB\,(no \lowLEE) + BAO.  

Setting aside the possibility of systematic errors in the Planck 2018 \lowLEE\ data, there are analysis and modeling choices that affect the inferences of the same data for $\tau$.   
As we have seen in Sec.\ \ref{sec:data}, the SRoll2 analysis provides a higher central value for $\tau$ but with smaller errors. In the bottom panel of Fig \ref{fig:tau}, we can see that this relaxes the tension slightly to $2.5\sigma$.  
The RELIKE reionization analysis with SRoll2, which allows for variations from the step-like reionization form further relaxes the tension to $2.3\sigma$.  While these changes do not allow a full reconciliation, they do weaken the preference for lower than minimal mass neutrinos.  

To quantify this weakening in a way that can more easily be mapped to changes in the constraints on $\tau$, we consider these modifications to be priors on $\tau$ in the CMB\,(no \lowLEE) analysis. This general approach is useful for comparing to other types of (non-CMB) constraints on $\tau$, for example from direct high-$z$ observations (see Sec.~\ref{sec:discussion}) that may suggest similar shifts in the future.  

The one drawback to this approach is that the CMB prior from no \lowLEE\ actually reflects all $\tau$ constraints from 
the primary CMB, including the smoothing of the acoustic peaks.  While this is a small component of the constraint, if we add it back to the analysis of primary CMB anisotropies as a prior, we double-count this peak-smoothing information.

To quantify this double-counting, we compare the CMB\,(\lowLEE\ prior)+BAO analysis to the CMB\,+\,BAO analysis (which includes \lowLEE\ data by default). In Tab.~\ref{tab:chi2}, we see that double counting shifts the best fit $\tau$ from $0.065$ to $0.066$ and changes $\Delta\chi^2$ between $\mnu=0$ and $\mnu=0.06$ by $0.2$ and that of $\Delta\chisqnoEE$ by $0.4$. 
Since the preference for higher $\tau$ is driven by lensing reconstruction and BAO, as this small shift shows, we can interpret the impact of analysis changes while neglecting this small double-counting. We reserve a full analysis for future work.

With just SRoll2, the combination of higher $\tau$ and smaller error bars leaves only a small effect on the neutrino mass tension as quantified in Tab.~\ref{tab:chi2}.
The RELIKE model has a larger impact 
and brings CMB\,(RELIKE)+BAO to
\begin{equation}
    \Delta\chi^2(\mnu=0) -\Delta\chi^2(0.06)\approx -3.85 \,,
    \label{eq:cmb_relike_bao}
\end{equation}
and allows an improvement of $\Delta\chisqnoEE=-1.7$ over the fiducial model.   The posterior shift in $\tau$ is a more substantial change from $\tau = 0.065$ to $0.07$ for the best-fit.  

\begin{figure}
    \centering
    \includegraphics[width = 0.95\linewidth]{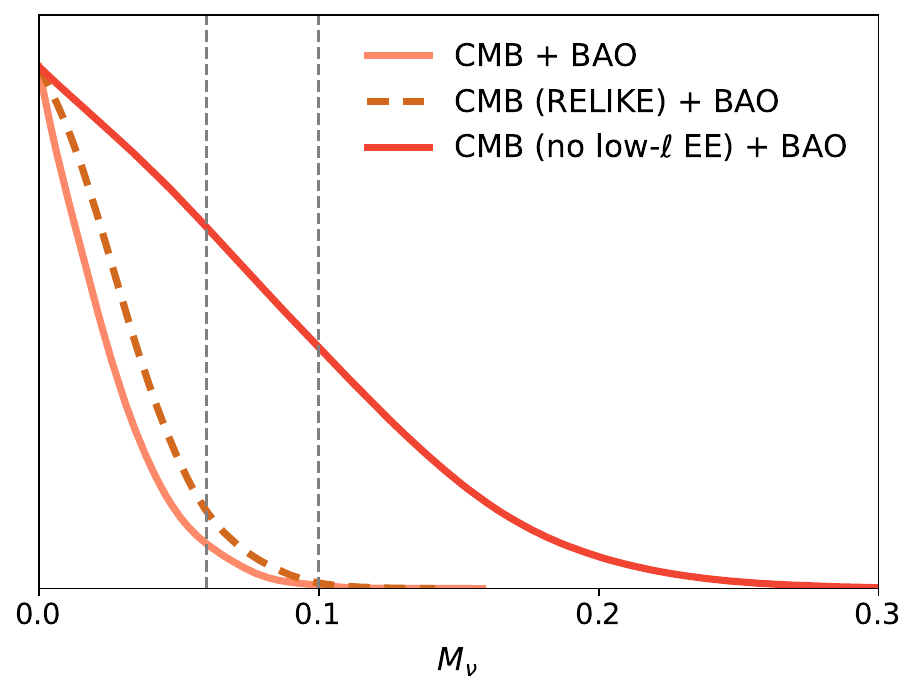}
    \vskip -0.3cm
    \caption{Posterior distributions for $\mnu$ with CMB\,+\,BAO vs CMB\,(RELIKE)+BAO and CMB\,(no \lowLEE)+BAO.  
    While the RELIKE $\tau$ prior only partially relieves tension with minimal neutrino masses (vertical dashed lines) when compared with no \lowLEE\ constraint, it improves the probability of $\mnu\ge 0.06$ by a factor of 1.35 and the exclusion falls below the 95\% CL.
    }
    \label{fig:mnu}
\end{figure}

While small, this changes the posterior probability $\mnu$ as shown in Fig.~\ref{fig:mnu}.  
In particular, the probability  of $\mnu\ge 0.06$ increases by a factor of $1.35$ and reduces the exclusion to the $93\%$ CL level, i.e.\ below the nominal $95\%$ CL of a tension.  

Apart from freedom in reionization models and analyses, there is also some leeway in inflationary models to raise $\tau$ further.  Beyond slow-roll inflation has the ability to do this by reducing the power ${\cal P}_{\cal R}$ in curvature fluctuations on large scales \cite{Mortonson:2009xk}, which is also mildly preferred by the large-angle anomalies in the TT spectrum. This can boost the central value of $\tau$ by $\sim 0.01$, and equally importantly, broaden its tail to high values \cite{Obied:2018qdr}.  

\begin{figure}
    \centering
    \includegraphics[width =0.95 \linewidth]{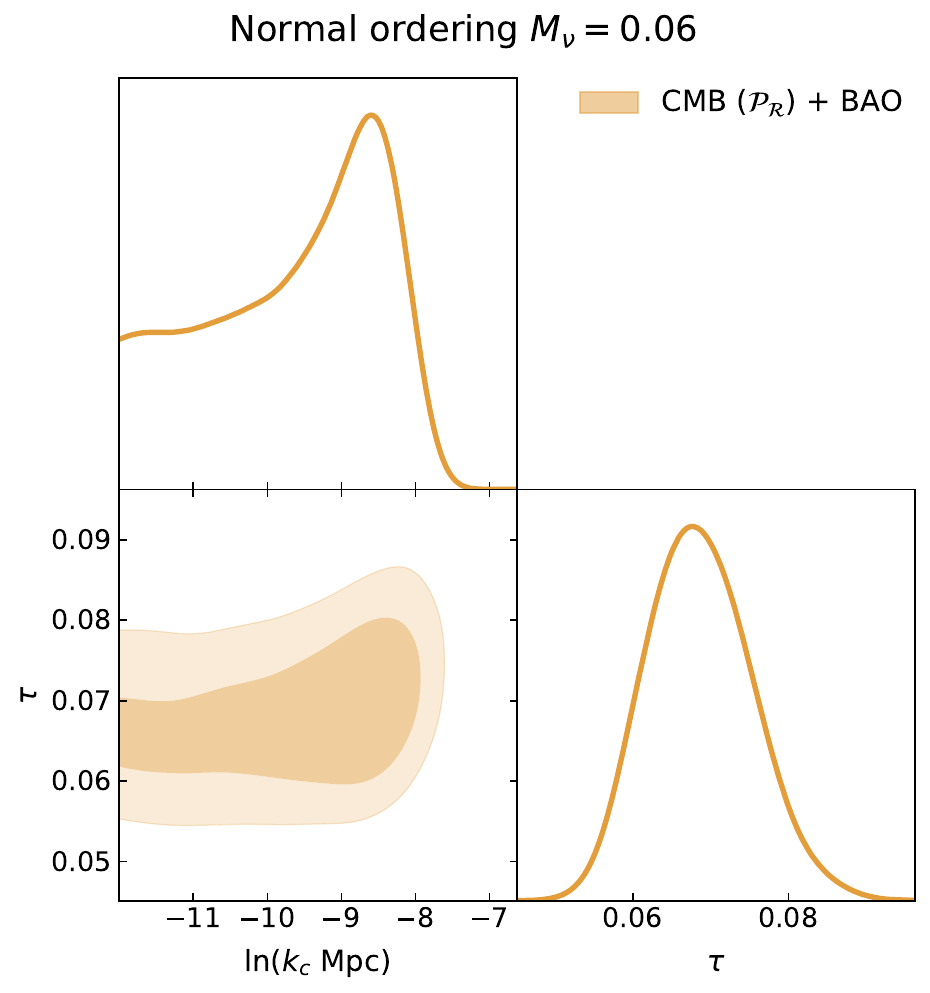}
    \vskip -0.2cm
    \caption{CMB\,(${\cal P}_{\cal R}$)\,+\,BAO with $\mnu=0.06$, where the curvature power spectrum is suppressed for $k\lesssim k_c$ (see Eq.~(\ref{eq:kc})) and the SRoll2 likelihood \cite{Delouis:2019bub} is used for the low-$\ell$ EE data. 
    Large-angle temperature anomalies prefer a value of
    $\ln (k_c \, {\rm Mpc})$ between $-8$ and $-9$, 
    where the low-$\ell$ polarization suppression allows larger $\tau$.
    }
\label{fig:trunc_Pk_tau}
\end{figure}

As a proof of principle for the impact of ${\cal P}_{\cal R}$\,, we analyze the 
exponential-cutoff model of Eq.~(\ref{eq:kc}) with SRoll2 data as
CMB\,(${\cal P}_{\cal R}$)\,+\,BAO in 
Fig.~\ref{fig:trunc_Pk_tau}.   Notice that the same preference from the low-$\ell$ TT power anomalies for a finite $\ln (k_c \, {\rm Mpc}) \sim -8.5$ also allows larger $\tau$ due to the suppression of \lowLEE\ power \cite{Mortonson:2009xk}.   In Tab.~\ref{tab:chi2},
we show the impact of of $k_c$ on the neutrino mass tension with CMB\,(${\cal P}_{\cal R}$)\,+\,BAO

\begin{equation}
    \Delta\chi^2(\mnu=0) -\Delta\chi^2(0.06)\approx -4.4 \,.
    \label{eq:Pk_chi2}
\end{equation}
This a slightly larger difference  than in Eq.\ (\ref{eq:cmb_relike_bao}). Conversely, the $\Delta\chisqnoEE$ between $\mnu=0$ and $\mnu = 0.06$ of CMB\,(${\cal P}_{\cal R}$)\,+\,BAO ($-2.5$) is slightly better than that of CMB\,(RELIKE) + BAO ($-2.8$). Therefore, while Eq. \eqref{eq:Pk_chi2} reflects a penalty from $\Delta \chi^2_\tau$\,, the difference mainly comes from the improvement of the fit at $\mnu=0$ rather than a degradation at $\mnu=0.06$.
Likewise, both models have a better $\Delta \chi^2$ and $\Delta\chisqnoEE$ than their non-$k_c$ alternatives in part due to the ability to adjust parameters to account for the low TT power anomaly.
 The consequence for the current tension is that $k_c$ has a comparable ability as reionization modeling for raising the best-fit optical depth, which becomes $\tau = 0.07$ for $\mnu=0.06$ and 0.064 for $\mnu=0$.   

The combination of reionization and inflaton model freedom could therefore double the change in $\tau$ and bring these values closer to the $\tau \sim 0.09$ preferred by the no \lowLEE\ case.   In fact, when optimized over any inflationary model, Ref.\ \cite{Obied:2018qdr} showed  that the two combined can essentially directly add their separate freedoms. 
While currently these improvements do not justify the added model complexity, should the tension become stronger with future data, using a combination of reionization and/or  even more complex inflationary model freedom could be favored. We leave these explorations for future work.

\section{Discussion}
\label{sec:discussion}

We have shown that the CMB\,+\,BAO tension with minimal-mass neutrinos in $\Lambda$CDM can be transferred to a tension in the optical depth $\tau$ at $\mnu=0.06$,  which raises the predicted amount of lensing in the CMB.  To fully resolve the tension with reionization, the optical depth 
would need to rise to $\tau = 0.091\pm 0.011$ from its \lowLEE\ and standard-reionization result of $\tau =0.0592\pm 0.0062$ with slow-roll inflation.  In $\Lambda$CDM, this shifts the geometric calibration of the sound horizon and consequently also raises $H_0$.
Beyond $\Lambda$CDM, shifting $\tau$ would also reduce the CMB\,+\,BAO preference for $w_0-w_a$ dynamical dark energy to below 95\% CL.

Employing the enhanced optical depth of the SRoll2 reanalysis and the extended reionization modeling of RELIKE only allows $\tau = 0.0619\pm 0.0062$ but does increase the posterior probability of $\mnu\ge 0.06$ by a factor of 1.35 which removes the 95\% CL exclusion.  Alternately, changing the initial curvature power spectrum in a manner previously suggested for large-angle temperature anomalies would allow a comparable effect on $\tau$.   The combination of reionization and inflation modeling  as in Ref.~\cite{Obied:2018qdr} could resolve the optical depth tension even further.  Finding a concrete model beyond single-field slow-roll inflation that does so would be interesting in the future.

\begin{figure}[t]
    \centering
    \vskip 0.2cm
    \includegraphics[width =0.95 \linewidth]{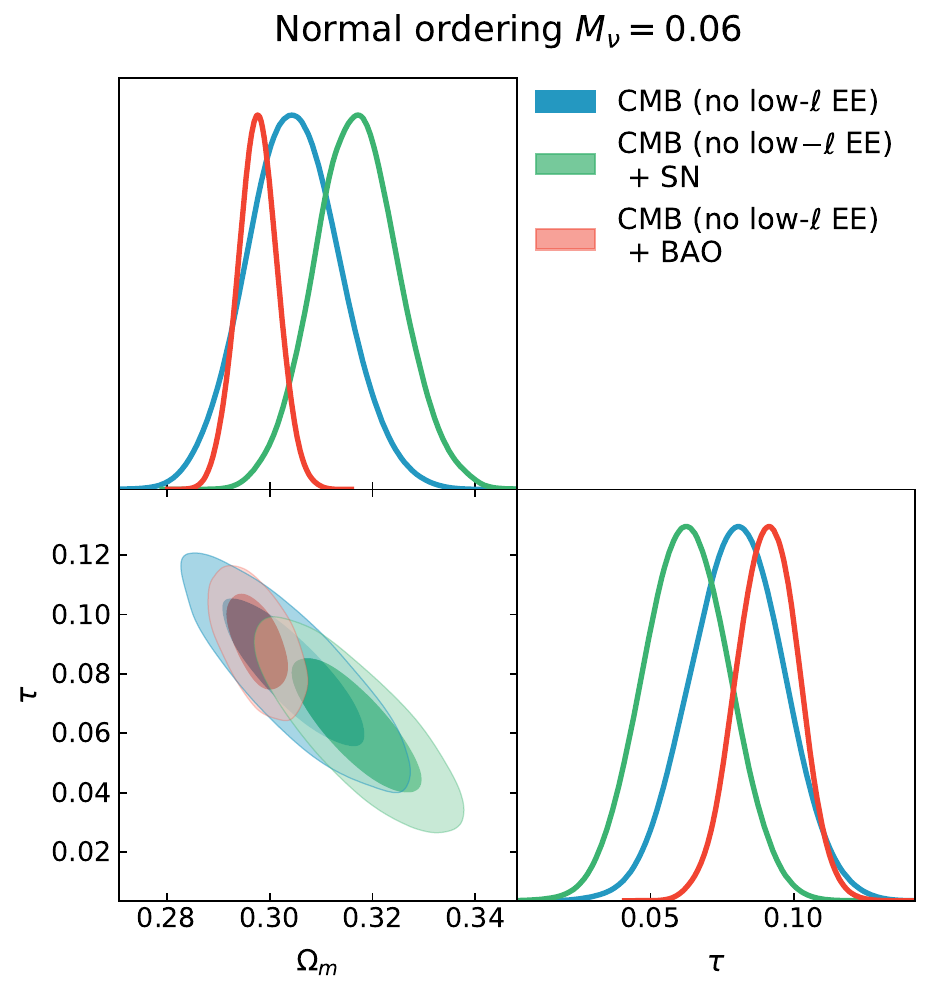}
    \vskip -0.2cm
    \caption{SN vs. BAO combinations with CMB\,(no lowL EE) for $\Omega_m-\tau$.  Removing the low-$\ell$ EE constraints in CMB can bring $\Omega_m$ and $\tau$ into consistency with either BAO or SN but not both.
    }
\label{fig:CMBBAOSN}
\end{figure}

In spite of this relaxation of the tension within $\Lambda$CDM, the optical-depth solution cannot simultaneously resolve the tension between CMB, BAO and SN that leads to a joint preference for dynamical dark energy, even if we altogether drop the \lowLEE\ constraint on $\tau$.  
While CMB\,(no \lowLEE) can accommodate the preferred range on $\Omega_m$ of either BAO or SN, the BAO and SN preferences are themselves in tension \cite{DESI:2025zgx,Tang:2024lmo}. 
We show this in Fig.~\ref{fig:CMBBAOSN} along with the correlated preference for high and low $\tau$ respectively.  
Tension between the three datasets is reduced in the $w_0-w_a$ dynamical dark energy space but only fully if the dark energy evolves across the phantom divide of $w=-1$, or with a more complicated dark sector which mimics that expansion history \cite{DESI:2025fii}.
This resolution also relaxes the tension with minimal-mass neutrinos \cite{DESI:2025ejh}. 

Given the theoretical problems and implications of such a phantom solution, it is still important to examine the weak links in this chain of inferences as we have done here with $\tau$ for CMB\,+\,BAO.   
Upcoming measurements of $\tau$ will help clarify the compatibility of $\Lambda$CDM with these datasets.   
Using the CMB, the ground-based Cosmology Large Angular Scale Surveyor (CLASS) telescope \cite{Harrington:2016jrz} aims to measure $\tau$  to near cosmic variance precision \cite{Watts:2018etg}, and has already reported measurements in cross correlation with Planck \cite{CLASS:2025khf}. Further in the future, the proposed satellite LiteBIRD \cite{LiteBIRD:2020khw} could measure the principal components of the ionization history itself
\cite{Sakamoto:2022nth}.

Beyond the CMB, direct high-redshift observations may prove incisive on these tensions.
For example, within standard models of reionization, 21cm measurements are projected to improve
constraints on $\tau$ substantially and break the CMB degeneracy with $\mnu$ (e.g.~\cite{Liu:2015txa,Shmueli:2023box}).
Conversely, measurements of the suppression of structure below the free streaming scale will isolate the neutrino effects (e.g.~\cite{DESI:2025ejh} and references therein).
In fact, JWST measurements of high-redshift galaxies 
\cite{2024ApJ...969L...2F,2022ApJ...938L..15C,2023arXiv230602465E,2023ApJS..265....5H} may  already be providing hints (e.g.~\cite{Munoz:2024fas}) for higher $\tau$ and earlier reionization. 

\smallskip

{\it Note added:} After completion of this work, Ref.~\cite{Sailer:2025lxj} appeared which reaches related conclusions about removing optical depth constraints.
\vfill

\acknowledgements We thank Tom Crawford for extensive help throughout the project and direct development of its initial stages. We thank Fei Ge for help with SPT-3G MUSE and Laura Herold, Leah Jenks, Austin Joyce, Harley Katz, Ely Kovetz, Adam Lidz for useful discussions. This work was supported by U.S.\ Dept.\ of Energy contract DE-FG02-13ER41958 and the Simons Foundation. TJ acknowledges additional funding from National Science Foundation Award OPP-1852617; TK  by the Kavli Institute for Cosmological Physics at the University of Chicago through an endowment from the Kavli Foundation. 
Computing resources were provided by the  University of Chicago  Research
Computing Center through the Kavli Institute for Cosmological Physics.

\vfill

\bibliographystyle{apsrev4-2}
\bibliography{nutau}

\begin{thebibliography}{64}%
\makeatletter
\providecommand \@ifxundefined [1]{%
 \@ifx{#1\undefined}
}%
\providecommand \@ifnum [1]{%
 \ifnum #1\expandafter \@firstoftwo
 \else \expandafter \@secondoftwo
 \fi
}%
\providecommand \@ifx [1]{%
 \ifx #1\expandafter \@firstoftwo
 \else \expandafter \@secondoftwo
 \fi
}%
\providecommand \natexlab [1]{#1}%
\providecommand \enquote  [1]{``#1''}%
\providecommand \bibnamefont  [1]{#1}%
\providecommand \bibfnamefont [1]{#1}%
\providecommand \citenamefont [1]{#1}%
\providecommand \href@noop [0]{\@secondoftwo}%
\providecommand \href [0]{\begingroup \@sanitize@url \@href}%
\providecommand \@href[1]{\@@startlink{#1}\@@href}%
\providecommand \@@href[1]{\endgroup#1\@@endlink}%
\providecommand \@sanitize@url [0]{\catcode `\\12\catcode `\$12\catcode `\&12\catcode `\#12\catcode `\^12\catcode `\_12\catcode `\%12\relax}%
\providecommand \@@startlink[1]{}%
\providecommand \@@endlink[0]{}%
\providecommand \url  [0]{\begingroup\@sanitize@url \@url }%
\providecommand \@url [1]{\endgroup\@href {#1}{\urlprefix }}%
\providecommand \urlprefix  [0]{URL }%
\providecommand \Eprint [0]{\href }%
\providecommand \doibase [0]{https://doi.org/}%
\providecommand \selectlanguage [0]{\@gobble}%
\providecommand \bibinfo  [0]{\@secondoftwo}%
\providecommand \bibfield  [0]{\@secondoftwo}%
\providecommand \translation [1]{[#1]}%
\providecommand \BibitemOpen [0]{}%
\providecommand \bibitemStop [0]{}%
\providecommand \bibitemNoStop [0]{.\EOS\space}%
\providecommand \EOS [0]{\spacefactor3000\relax}%
\providecommand \BibitemShut  [1]{\csname bibitem#1\endcsname}%
\let\auto@bib@innerbib\@empty
\bibitem [{\citenamefont {Navas}\ \emph {et~al.}(2024)\citenamefont {Navas} \emph {et~al.}}]{ParticleDataGroup:2024cfk}%
  \BibitemOpen
  \bibfield  {author} {\bibinfo {author} {\bibfnamefont {S.}~\bibnamefont {Navas}} \emph {et~al.} (\bibinfo {collaboration} {Particle Data Group}),\ }\href {https://doi.org/10.1103/PhysRevD.110.030001} {\bibfield  {journal} {\bibinfo  {journal} {Phys. Rev. D}\ }\textbf {\bibinfo {volume} {110}},\ \bibinfo {pages} {030001} (\bibinfo {year} {2024})}\BibitemShut {NoStop}%
\bibitem [{\citenamefont {Di~Valentino}\ \emph {et~al.}(2024)\citenamefont {Di~Valentino}, \citenamefont {Gariazzo},\ and\ \citenamefont {Mena}}]{DiValentino:2024xsv}%
  \BibitemOpen
  \bibfield  {author} {\bibinfo {author} {\bibfnamefont {E.}~\bibnamefont {Di~Valentino}}, \bibinfo {author} {\bibfnamefont {S.}~\bibnamefont {Gariazzo}},\ and\ \bibinfo {author} {\bibfnamefont {O.}~\bibnamefont {Mena}},\ }\href@noop {} {\  (\bibinfo {year} {2024})},\ \Eprint {https://arxiv.org/abs/2404.19322} {arXiv:2404.19322 [astro-ph.CO]} \BibitemShut {NoStop}%
\bibitem [{\citenamefont {Esteban}\ \emph {et~al.}(2024)\citenamefont {Esteban}, \citenamefont {Gonzalez-Garcia}, \citenamefont {Maltoni}, \citenamefont {Martinez-Soler}, \citenamefont {Pinheiro},\ and\ \citenamefont {Schwetz}}]{Esteban:2024eli}%
  \BibitemOpen
  \bibfield  {author} {\bibinfo {author} {\bibfnamefont {I.}~\bibnamefont {Esteban}}, \bibinfo {author} {\bibfnamefont {M.~C.}\ \bibnamefont {Gonzalez-Garcia}}, \bibinfo {author} {\bibfnamefont {M.}~\bibnamefont {Maltoni}}, \bibinfo {author} {\bibfnamefont {I.}~\bibnamefont {Martinez-Soler}}, \bibinfo {author} {\bibfnamefont {J.~a.~P.}\ \bibnamefont {Pinheiro}},\ and\ \bibinfo {author} {\bibfnamefont {T.}~\bibnamefont {Schwetz}},\ }\href {https://doi.org/10.1007/JHEP12(2024)216} {\bibfield  {journal} {\bibinfo  {journal} {JHEP}\ }\textbf {\bibinfo {volume} {12}},\ \bibinfo {pages} {216}},\ \Eprint {https://arxiv.org/abs/2410.05380} {arXiv:2410.05380 [hep-ph]} \BibitemShut {NoStop}%
\bibitem [{\citenamefont {Adame}\ \emph {et~al.}(2024)\citenamefont {Adame} \emph {et~al.}}]{DESI:2024mwx}%
  \BibitemOpen
  \bibfield  {author} {\bibinfo {author} {\bibfnamefont {A.~G.}\ \bibnamefont {Adame}} \emph {et~al.} (\bibinfo {collaboration} {DESI}),\ }\href@noop {} {\  (\bibinfo {year} {2024})},\ \Eprint {https://arxiv.org/abs/2404.03002} {arXiv:2404.03002 [astro-ph.CO]} \BibitemShut {NoStop}%
\bibitem [{\citenamefont {Craig}\ \emph {et~al.}(2024)\citenamefont {Craig}, \citenamefont {Green}, \citenamefont {Meyers},\ and\ \citenamefont {Rajendran}}]{Craig:2024tky}%
  \BibitemOpen
  \bibfield  {author} {\bibinfo {author} {\bibfnamefont {N.}~\bibnamefont {Craig}}, \bibinfo {author} {\bibfnamefont {D.}~\bibnamefont {Green}}, \bibinfo {author} {\bibfnamefont {J.}~\bibnamefont {Meyers}},\ and\ \bibinfo {author} {\bibfnamefont {S.}~\bibnamefont {Rajendran}},\ }\href {https://doi.org/10.1007/JHEP09(2024)097} {\bibfield  {journal} {\bibinfo  {journal} {JHEP}\ }\textbf {\bibinfo {volume} {09}},\ \bibinfo {pages} {097}},\ \Eprint {https://arxiv.org/abs/2405.00836} {arXiv:2405.00836 [astro-ph.CO]} \BibitemShut {NoStop}%
\bibitem [{\citenamefont {Green}\ and\ \citenamefont {Meyers}(2024)}]{Green:2024xbb}%
  \BibitemOpen
  \bibfield  {author} {\bibinfo {author} {\bibfnamefont {D.}~\bibnamefont {Green}}\ and\ \bibinfo {author} {\bibfnamefont {J.}~\bibnamefont {Meyers}},\ }\href@noop {} {\  (\bibinfo {year} {2024})},\ \Eprint {https://arxiv.org/abs/2407.07878} {arXiv:2407.07878 [astro-ph.CO]} \BibitemShut {NoStop}%
\bibitem [{\citenamefont {Loverde}\ and\ \citenamefont {Weiner}(2024)}]{Loverde:2024nfi}%
  \BibitemOpen
  \bibfield  {author} {\bibinfo {author} {\bibfnamefont {M.}~\bibnamefont {Loverde}}\ and\ \bibinfo {author} {\bibfnamefont {Z.~J.}\ \bibnamefont {Weiner}},\ }\href {https://doi.org/10.1088/1475-7516/2024/12/048} {\bibfield  {journal} {\bibinfo  {journal} {JCAP}\ }\textbf {\bibinfo {volume} {12}},\ \bibinfo {pages} {048}},\ \Eprint {https://arxiv.org/abs/2410.00090} {arXiv:2410.00090 [astro-ph.CO]} \BibitemShut {NoStop}%
\bibitem [{\citenamefont {Elbers}\ \emph {et~al.}(2025)\citenamefont {Elbers} \emph {et~al.}}]{DESI:2025ejh}%
  \BibitemOpen
  \bibfield  {author} {\bibinfo {author} {\bibfnamefont {W.}~\bibnamefont {Elbers}} \emph {et~al.} (\bibinfo {collaboration} {DESI}),\ }\href@noop {} {\  (\bibinfo {year} {2025})},\ \Eprint {https://arxiv.org/abs/2503.14744} {arXiv:2503.14744 [astro-ph.CO]} \BibitemShut {NoStop}%
\bibitem [{\citenamefont {Lynch}\ and\ \citenamefont {Knox}(2025)}]{Lynch:2025ine}%
  \BibitemOpen
  \bibfield  {author} {\bibinfo {author} {\bibfnamefont {G.~P.}\ \bibnamefont {Lynch}}\ and\ \bibinfo {author} {\bibfnamefont {L.}~\bibnamefont {Knox}},\ }\href@noop {} {\  (\bibinfo {year} {2025})},\ \Eprint {https://arxiv.org/abs/2503.14470} {arXiv:2503.14470 [astro-ph.CO]} \BibitemShut {NoStop}%
\bibitem [{\citenamefont {Naredo-Tuero}\ \emph {et~al.}(2024)\citenamefont {Naredo-Tuero}, \citenamefont {Escudero}, \citenamefont {Fern\'andez-Mart\'\i{}nez}, \citenamefont {Marcano},\ and\ \citenamefont {Poulin}}]{Naredo-Tuero:2024sgf}%
  \BibitemOpen
  \bibfield  {author} {\bibinfo {author} {\bibfnamefont {D.}~\bibnamefont {Naredo-Tuero}}, \bibinfo {author} {\bibfnamefont {M.}~\bibnamefont {Escudero}}, \bibinfo {author} {\bibfnamefont {E.}~\bibnamefont {Fern\'andez-Mart\'\i{}nez}}, \bibinfo {author} {\bibfnamefont {X.}~\bibnamefont {Marcano}},\ and\ \bibinfo {author} {\bibfnamefont {V.}~\bibnamefont {Poulin}},\ }\href {https://doi.org/10.1103/PhysRevD.110.123537} {\bibfield  {journal} {\bibinfo  {journal} {Phys. Rev. D}\ }\textbf {\bibinfo {volume} {110}},\ \bibinfo {pages} {123537} (\bibinfo {year} {2024})},\ \Eprint {https://arxiv.org/abs/2407.13831} {arXiv:2407.13831 [astro-ph.CO]} \BibitemShut {NoStop}%
\bibitem [{\citenamefont {Roy~Choudhury}\ and\ \citenamefont {Okumura}(2024)}]{RoyChoudhury:2024wri}%
  \BibitemOpen
  \bibfield  {author} {\bibinfo {author} {\bibfnamefont {S.}~\bibnamefont {Roy~Choudhury}}\ and\ \bibinfo {author} {\bibfnamefont {T.}~\bibnamefont {Okumura}},\ }\href {https://doi.org/10.3847/2041-8213/ad8c26} {\bibfield  {journal} {\bibinfo  {journal} {Astrophys. J. Lett.}\ }\textbf {\bibinfo {volume} {976}},\ \bibinfo {pages} {L11} (\bibinfo {year} {2024})},\ \Eprint {https://arxiv.org/abs/2409.13022} {arXiv:2409.13022 [astro-ph.CO]} \BibitemShut {NoStop}%
\bibitem [{\citenamefont {Rebou\c{c}as}\ \emph {et~al.}(2025)\citenamefont {Rebou\c{c}as}, \citenamefont {de~Souza}, \citenamefont {Zhong}, \citenamefont {Miranda},\ and\ \citenamefont {Rosenfeld}}]{Reboucas:2024smm}%
  \BibitemOpen
  \bibfield  {author} {\bibinfo {author} {\bibfnamefont {J.}~\bibnamefont {Rebou\c{c}as}}, \bibinfo {author} {\bibfnamefont {D.~H.~F.}\ \bibnamefont {de~Souza}}, \bibinfo {author} {\bibfnamefont {K.}~\bibnamefont {Zhong}}, \bibinfo {author} {\bibfnamefont {V.}~\bibnamefont {Miranda}},\ and\ \bibinfo {author} {\bibfnamefont {R.}~\bibnamefont {Rosenfeld}},\ }\href {https://doi.org/10.1088/1475-7516/2025/02/024} {\bibfield  {journal} {\bibinfo  {journal} {JCAP}\ }\textbf {\bibinfo {volume} {02}},\ \bibinfo {pages} {024}},\ \Eprint {https://arxiv.org/abs/2408.14628} {arXiv:2408.14628 [astro-ph.CO]} \BibitemShut {NoStop}%
\bibitem [{\citenamefont {Jiang}\ \emph {et~al.}(2025)\citenamefont {Jiang}, \citenamefont {Giar\`e}, \citenamefont {Gariazzo}, \citenamefont {Dainotti}, \citenamefont {Di~Valentino}, \citenamefont {Mena}, \citenamefont {Pedrotti}, \citenamefont {da~Costa},\ and\ \citenamefont {Vagnozzi}}]{Jiang:2024viw}%
  \BibitemOpen
  \bibfield  {author} {\bibinfo {author} {\bibfnamefont {J.-Q.}\ \bibnamefont {Jiang}}, \bibinfo {author} {\bibfnamefont {W.}~\bibnamefont {Giar\`e}}, \bibinfo {author} {\bibfnamefont {S.}~\bibnamefont {Gariazzo}}, \bibinfo {author} {\bibfnamefont {M.~G.}\ \bibnamefont {Dainotti}}, \bibinfo {author} {\bibfnamefont {E.}~\bibnamefont {Di~Valentino}}, \bibinfo {author} {\bibfnamefont {O.}~\bibnamefont {Mena}}, \bibinfo {author} {\bibfnamefont {D.}~\bibnamefont {Pedrotti}}, \bibinfo {author} {\bibfnamefont {S.~S.}\ \bibnamefont {da~Costa}},\ and\ \bibinfo {author} {\bibfnamefont {S.}~\bibnamefont {Vagnozzi}},\ }\href {https://doi.org/10.1088/1475-7516/2025/01/153} {\bibfield  {journal} {\bibinfo  {journal} {JCAP}\ }\textbf {\bibinfo {volume} {01}},\ \bibinfo {pages} {153}},\ \Eprint {https://arxiv.org/abs/2407.18047} {arXiv:2407.18047 [astro-ph.CO]} \BibitemShut {NoStop}%
\bibitem [{\citenamefont {Tang}\ \emph {et~al.}(2025)\citenamefont {Tang}, \citenamefont {Brout}, \citenamefont {Karwal}, \citenamefont {Chang}, \citenamefont {Miranda},\ and\ \citenamefont {Vincenzi}}]{Tang:2024lmo}%
  \BibitemOpen
  \bibfield  {author} {\bibinfo {author} {\bibfnamefont {X.~T.}\ \bibnamefont {Tang}}, \bibinfo {author} {\bibfnamefont {D.}~\bibnamefont {Brout}}, \bibinfo {author} {\bibfnamefont {T.}~\bibnamefont {Karwal}}, \bibinfo {author} {\bibfnamefont {C.}~\bibnamefont {Chang}}, \bibinfo {author} {\bibfnamefont {V.}~\bibnamefont {Miranda}},\ and\ \bibinfo {author} {\bibfnamefont {M.}~\bibnamefont {Vincenzi}},\ }\href {https://doi.org/10.3847/2041-8213/adc4da} {\bibfield  {journal} {\bibinfo  {journal} {Astrophys. J. Lett.}\ }\textbf {\bibinfo {volume} {983}},\ \bibinfo {pages} {L27} (\bibinfo {year} {2025})},\ \Eprint {https://arxiv.org/abs/2412.04430} {arXiv:2412.04430 [astro-ph.CO]} \BibitemShut {NoStop}%
\bibitem [{\citenamefont {Calabrese}\ \emph {et~al.}(2008)\citenamefont {Calabrese}, \citenamefont {Slosar}, \citenamefont {Melchiorri}, \citenamefont {Smoot},\ and\ \citenamefont {Zahn}}]{Calabrese:2008rt}%
  \BibitemOpen
  \bibfield  {author} {\bibinfo {author} {\bibfnamefont {E.}~\bibnamefont {Calabrese}}, \bibinfo {author} {\bibfnamefont {A.}~\bibnamefont {Slosar}}, \bibinfo {author} {\bibfnamefont {A.}~\bibnamefont {Melchiorri}}, \bibinfo {author} {\bibfnamefont {G.~F.}\ \bibnamefont {Smoot}},\ and\ \bibinfo {author} {\bibfnamefont {O.}~\bibnamefont {Zahn}},\ }\href {https://doi.org/10.1103/PhysRevD.77.123531} {\bibfield  {journal} {\bibinfo  {journal} {Phys. Rev. D}\ }\textbf {\bibinfo {volume} {77}},\ \bibinfo {pages} {123531} (\bibinfo {year} {2008})},\ \Eprint {https://arxiv.org/abs/0803.2309} {arXiv:0803.2309 [astro-ph]} \BibitemShut {NoStop}%
\bibitem [{\citenamefont {Ge}\ \emph {et~al.}(2024)\citenamefont {Ge} \emph {et~al.}}]{SPT-3G:2024atg}%
  \BibitemOpen
  \bibfield  {author} {\bibinfo {author} {\bibfnamefont {F.}~\bibnamefont {Ge}} \emph {et~al.} (\bibinfo {collaboration} {SPT-3G}),\ }\href@noop {} {\  (\bibinfo {year} {2024})},\ \Eprint {https://arxiv.org/abs/2411.06000} {arXiv:2411.06000 [astro-ph.CO]} \BibitemShut {NoStop}%
\bibitem [{\citenamefont {Smith}\ \emph {et~al.}(2006)\citenamefont {Smith}, \citenamefont {Hu},\ and\ \citenamefont {Kaplinghat}}]{Smith:2006nk}%
  \BibitemOpen
  \bibfield  {author} {\bibinfo {author} {\bibfnamefont {K.~M.}\ \bibnamefont {Smith}}, \bibinfo {author} {\bibfnamefont {W.}~\bibnamefont {Hu}},\ and\ \bibinfo {author} {\bibfnamefont {M.}~\bibnamefont {Kaplinghat}},\ }\href {https://doi.org/10.1103/PhysRevD.74.123002} {\bibfield  {journal} {\bibinfo  {journal} {Phys. Rev. D}\ }\textbf {\bibinfo {volume} {74}},\ \bibinfo {pages} {123002} (\bibinfo {year} {2006})},\ \Eprint {https://arxiv.org/abs/astro-ph/0607315} {arXiv:astro-ph/0607315} \BibitemShut {NoStop}%
\bibitem [{\citenamefont {Allison}\ \emph {et~al.}(2015)\citenamefont {Allison}, \citenamefont {Caucal}, \citenamefont {Calabrese}, \citenamefont {Dunkley},\ and\ \citenamefont {Louis}}]{Allison:2015qca}%
  \BibitemOpen
  \bibfield  {author} {\bibinfo {author} {\bibfnamefont {R.}~\bibnamefont {Allison}}, \bibinfo {author} {\bibfnamefont {P.}~\bibnamefont {Caucal}}, \bibinfo {author} {\bibfnamefont {E.}~\bibnamefont {Calabrese}}, \bibinfo {author} {\bibfnamefont {J.}~\bibnamefont {Dunkley}},\ and\ \bibinfo {author} {\bibfnamefont {T.}~\bibnamefont {Louis}},\ }\href {https://doi.org/10.1103/PhysRevD.92.123535} {\bibfield  {journal} {\bibinfo  {journal} {Phys. Rev. D}\ }\textbf {\bibinfo {volume} {92}},\ \bibinfo {pages} {123535} (\bibinfo {year} {2015})},\ \Eprint {https://arxiv.org/abs/1509.07471} {arXiv:1509.07471 [astro-ph.CO]} \BibitemShut {NoStop}%
\bibitem [{\citenamefont {Kogut}\ \emph {et~al.}(2003)\citenamefont {Kogut} \emph {et~al.}}]{WMAP:2003ggs}%
  \BibitemOpen
  \bibfield  {author} {\bibinfo {author} {\bibfnamefont {A.}~\bibnamefont {Kogut}} \emph {et~al.} (\bibinfo {collaboration} {WMAP}),\ }\href {https://doi.org/10.1086/377219} {\bibfield  {journal} {\bibinfo  {journal} {Astrophys. J. Suppl.}\ }\textbf {\bibinfo {volume} {148}},\ \bibinfo {pages} {161} (\bibinfo {year} {2003})},\ \Eprint {https://arxiv.org/abs/astro-ph/0302213} {arXiv:astro-ph/0302213} \BibitemShut {NoStop}%
\bibitem [{\citenamefont {Hinshaw}\ \emph {et~al.}(2013)\citenamefont {Hinshaw} \emph {et~al.}}]{WMAP:2012nax}%
  \BibitemOpen
  \bibfield  {author} {\bibinfo {author} {\bibfnamefont {G.}~\bibnamefont {Hinshaw}} \emph {et~al.} (\bibinfo {collaboration} {WMAP}),\ }\href {https://doi.org/10.1088/0067-0049/208/2/19} {\bibfield  {journal} {\bibinfo  {journal} {Astrophys. J. Suppl.}\ }\textbf {\bibinfo {volume} {208}},\ \bibinfo {pages} {19} (\bibinfo {year} {2013})},\ \Eprint {https://arxiv.org/abs/1212.5226} {arXiv:1212.5226 [astro-ph.CO]} \BibitemShut {NoStop}%
\bibitem [{\citenamefont {Ade}\ \emph {et~al.}(2014{\natexlab{a}})\citenamefont {Ade} \emph {et~al.}}]{Planck:2013pxb}%
  \BibitemOpen
  \bibfield  {author} {\bibinfo {author} {\bibfnamefont {P.~A.~R.}\ \bibnamefont {Ade}} \emph {et~al.} (\bibinfo {collaboration} {Planck}),\ }\href {https://doi.org/10.1051/0004-6361/201321591} {\bibfield  {journal} {\bibinfo  {journal} {Astron. Astrophys.}\ }\textbf {\bibinfo {volume} {571}},\ \bibinfo {pages} {A16} (\bibinfo {year} {2014}{\natexlab{a}})},\ \Eprint {https://arxiv.org/abs/1303.5076} {arXiv:1303.5076 [astro-ph.CO]} \BibitemShut {NoStop}%
\bibitem [{\citenamefont {Ade}\ \emph {et~al.}(2016)\citenamefont {Ade} \emph {et~al.}}]{Planck:2015fie}%
  \BibitemOpen
  \bibfield  {author} {\bibinfo {author} {\bibfnamefont {P.~A.~R.}\ \bibnamefont {Ade}} \emph {et~al.} (\bibinfo {collaboration} {Planck}),\ }\href {https://doi.org/10.1051/0004-6361/201525830} {\bibfield  {journal} {\bibinfo  {journal} {Astron. Astrophys.}\ }\textbf {\bibinfo {volume} {594}},\ \bibinfo {pages} {A13} (\bibinfo {year} {2016})},\ \Eprint {https://arxiv.org/abs/1502.01589} {arXiv:1502.01589 [astro-ph.CO]} \BibitemShut {NoStop}%
\bibitem [{\citenamefont {Aghanim}\ \emph {et~al.}(2020)\citenamefont {Aghanim} \emph {et~al.}}]{Planck:2018vyg}%
  \BibitemOpen
  \bibfield  {author} {\bibinfo {author} {\bibfnamefont {N.}~\bibnamefont {Aghanim}} \emph {et~al.} (\bibinfo {collaboration} {Planck}),\ }\href {https://doi.org/10.1051/0004-6361/201833910} {\bibfield  {journal} {\bibinfo  {journal} {Astron. Astrophys.}\ }\textbf {\bibinfo {volume} {641}},\ \bibinfo {pages} {A6} (\bibinfo {year} {2020})},\ \bibinfo {note} {[Erratum: Astron.Astrophys. 652, C4 (2021)]},\ \Eprint {https://arxiv.org/abs/1807.06209} {arXiv:1807.06209 [astro-ph.CO]} \BibitemShut {NoStop}%
\bibitem [{\citenamefont {Hu}\ and\ \citenamefont {Holder}(2003)}]{Hu:2003gh}%
  \BibitemOpen
  \bibfield  {author} {\bibinfo {author} {\bibfnamefont {W.}~\bibnamefont {Hu}}\ and\ \bibinfo {author} {\bibfnamefont {G.~P.}\ \bibnamefont {Holder}},\ }\href {https://doi.org/10.1103/PhysRevD.68.023001} {\bibfield  {journal} {\bibinfo  {journal} {Phys. Rev. D}\ }\textbf {\bibinfo {volume} {68}},\ \bibinfo {pages} {023001} (\bibinfo {year} {2003})},\ \Eprint {https://arxiv.org/abs/astro-ph/0303400} {arXiv:astro-ph/0303400} \BibitemShut {NoStop}%
\bibitem [{\citenamefont {Heinrich}\ \emph {et~al.}(2017)\citenamefont {Heinrich}, \citenamefont {Miranda},\ and\ \citenamefont {Hu}}]{Heinrich:2016ojb}%
  \BibitemOpen
  \bibfield  {author} {\bibinfo {author} {\bibfnamefont {C.~H.}\ \bibnamefont {Heinrich}}, \bibinfo {author} {\bibfnamefont {V.}~\bibnamefont {Miranda}},\ and\ \bibinfo {author} {\bibfnamefont {W.}~\bibnamefont {Hu}},\ }\href {https://doi.org/10.1103/PhysRevD.95.023513} {\bibfield  {journal} {\bibinfo  {journal} {Phys. Rev. D}\ }\textbf {\bibinfo {volume} {95}},\ \bibinfo {pages} {023513} (\bibinfo {year} {2017})},\ \Eprint {https://arxiv.org/abs/1609.04788} {arXiv:1609.04788 [astro-ph.CO]} \BibitemShut {NoStop}%
\bibitem [{\citenamefont {Delouis}\ \emph {et~al.}(2019)\citenamefont {Delouis}, \citenamefont {Pagano}, \citenamefont {Mottet}, \citenamefont {Puget},\ and\ \citenamefont {Vibert}}]{Delouis:2019bub}%
  \BibitemOpen
  \bibfield  {author} {\bibinfo {author} {\bibfnamefont {J.~M.}\ \bibnamefont {Delouis}}, \bibinfo {author} {\bibfnamefont {L.}~\bibnamefont {Pagano}}, \bibinfo {author} {\bibfnamefont {S.}~\bibnamefont {Mottet}}, \bibinfo {author} {\bibfnamefont {J.~L.}\ \bibnamefont {Puget}},\ and\ \bibinfo {author} {\bibfnamefont {L.}~\bibnamefont {Vibert}},\ }\href {https://doi.org/10.1051/0004-6361/201834882} {\bibfield  {journal} {\bibinfo  {journal} {Astron. Astrophys.}\ }\textbf {\bibinfo {volume} {629}},\ \bibinfo {pages} {A38} (\bibinfo {year} {2019})},\ \Eprint {https://arxiv.org/abs/1901.11386} {arXiv:1901.11386 [astro-ph.CO]} \BibitemShut {NoStop}%
\bibitem [{\citenamefont {Heinrich}\ and\ \citenamefont {Hu}(2021)}]{Heinrich:2021ufa}%
  \BibitemOpen
  \bibfield  {author} {\bibinfo {author} {\bibfnamefont {C.}~\bibnamefont {Heinrich}}\ and\ \bibinfo {author} {\bibfnamefont {W.}~\bibnamefont {Hu}},\ }\href {https://doi.org/10.1103/PhysRevD.104.063505} {\bibfield  {journal} {\bibinfo  {journal} {Phys. Rev. D}\ }\textbf {\bibinfo {volume} {104}},\ \bibinfo {pages} {063505} (\bibinfo {year} {2021})},\ \Eprint {https://arxiv.org/abs/2104.13998} {arXiv:2104.13998 [astro-ph.CO]} \BibitemShut {NoStop}%
\bibitem [{\citenamefont {Obied}\ \emph {et~al.}(2018)\citenamefont {Obied}, \citenamefont {Dvorkin}, \citenamefont {Heinrich}, \citenamefont {Hu},\ and\ \citenamefont {Miranda}}]{Obied:2018qdr}%
  \BibitemOpen
  \bibfield  {author} {\bibinfo {author} {\bibfnamefont {G.}~\bibnamefont {Obied}}, \bibinfo {author} {\bibfnamefont {C.}~\bibnamefont {Dvorkin}}, \bibinfo {author} {\bibfnamefont {C.}~\bibnamefont {Heinrich}}, \bibinfo {author} {\bibfnamefont {W.}~\bibnamefont {Hu}},\ and\ \bibinfo {author} {\bibfnamefont {V.}~\bibnamefont {Miranda}},\ }\href {https://doi.org/10.1103/PhysRevD.98.043518} {\bibfield  {journal} {\bibinfo  {journal} {Phys. Rev. D}\ }\textbf {\bibinfo {volume} {98}},\ \bibinfo {pages} {043518} (\bibinfo {year} {2018})},\ \Eprint {https://arxiv.org/abs/1803.01858} {arXiv:1803.01858 [astro-ph.CO]} \BibitemShut {NoStop}%
\bibitem [{\citenamefont {Carron}\ \emph {et~al.}(2022)\citenamefont {Carron}, \citenamefont {Mirmelstein},\ and\ \citenamefont {Lewis}}]{Carron:2022eyg}%
  \BibitemOpen
  \bibfield  {author} {\bibinfo {author} {\bibfnamefont {J.}~\bibnamefont {Carron}}, \bibinfo {author} {\bibfnamefont {M.}~\bibnamefont {Mirmelstein}},\ and\ \bibinfo {author} {\bibfnamefont {A.}~\bibnamefont {Lewis}},\ }\href {https://doi.org/10.1088/1475-7516/2022/09/039} {\bibfield  {journal} {\bibinfo  {journal} {JCAP}\ }\textbf {\bibinfo {volume} {09}},\ \bibinfo {pages} {039}},\ \Eprint {https://arxiv.org/abs/2206.07773} {arXiv:2206.07773 [astro-ph.CO]} \BibitemShut {NoStop}%
\bibitem [{\citenamefont {Qu}\ \emph {et~al.}(2024)\citenamefont {Qu} \emph {et~al.}}]{ACT:2023dou}%
  \BibitemOpen
  \bibfield  {author} {\bibinfo {author} {\bibfnamefont {F.~J.}\ \bibnamefont {Qu}} \emph {et~al.} (\bibinfo {collaboration} {ACT}),\ }\href {https://doi.org/10.3847/1538-4357/acfe06} {\bibfield  {journal} {\bibinfo  {journal} {Astrophys. J.}\ }\textbf {\bibinfo {volume} {962}},\ \bibinfo {pages} {112} (\bibinfo {year} {2024})},\ \Eprint {https://arxiv.org/abs/2304.05202} {arXiv:2304.05202 [astro-ph.CO]} \BibitemShut {NoStop}%
\bibitem [{\citenamefont {Qu}\ \emph {et~al.}(2025)\citenamefont {Qu} \emph {et~al.}}]{ACT:2025rvn}%
  \BibitemOpen
  \bibfield  {author} {\bibinfo {author} {\bibfnamefont {F.~J.}\ \bibnamefont {Qu}} \emph {et~al.} (\bibinfo {collaboration} {ACT, SPT-3G}),\ }\href@noop {} {\  (\bibinfo {year} {2025})},\ \Eprint {https://arxiv.org/abs/2504.20038} {arXiv:2504.20038 [astro-ph.CO]} \BibitemShut {NoStop}%
\bibitem [{\citenamefont {Abdul~Karim}\ \emph {et~al.}(2025)\citenamefont {Abdul~Karim} \emph {et~al.}}]{DESI:2025zgx}%
  \BibitemOpen
  \bibfield  {author} {\bibinfo {author} {\bibfnamefont {M.}~\bibnamefont {Abdul~Karim}} \emph {et~al.} (\bibinfo {collaboration} {DESI}),\ }\href@noop {} {\  (\bibinfo {year} {2025})},\ \Eprint {https://arxiv.org/abs/2503.14738} {arXiv:2503.14738 [astro-ph.CO]} \BibitemShut {NoStop}%
\bibitem [{\citenamefont {Abbott}\ \emph {et~al.}(2024)\citenamefont {Abbott} \emph {et~al.}}]{DES:2024jxu}%
  \BibitemOpen
  \bibfield  {author} {\bibinfo {author} {\bibfnamefont {T.~M.~C.}\ \bibnamefont {Abbott}} \emph {et~al.} (\bibinfo {collaboration} {DES}),\ }\href {https://doi.org/10.3847/2041-8213/ad6f9f} {\bibfield  {journal} {\bibinfo  {journal} {Astrophys. J. Lett.}\ }\textbf {\bibinfo {volume} {973}},\ \bibinfo {pages} {L14} (\bibinfo {year} {2024})},\ \Eprint {https://arxiv.org/abs/2401.02929} {arXiv:2401.02929 [astro-ph.CO]} \BibitemShut {NoStop}%
\bibitem [{\citenamefont {Hannestad}\ and\ \citenamefont {Schwetz}(2016)}]{Hannestad:2016fog}%
  \BibitemOpen
  \bibfield  {author} {\bibinfo {author} {\bibfnamefont {S.}~\bibnamefont {Hannestad}}\ and\ \bibinfo {author} {\bibfnamefont {T.}~\bibnamefont {Schwetz}},\ }\href {https://doi.org/10.1088/1475-7516/2016/11/035} {\bibfield  {journal} {\bibinfo  {journal} {JCAP}\ }\textbf {\bibinfo {volume} {11}},\ \bibinfo {pages} {035}},\ \Eprint {https://arxiv.org/abs/1606.04691} {arXiv:1606.04691 [astro-ph.CO]} \BibitemShut {NoStop}%
\bibitem [{\citenamefont {Herold}\ and\ \citenamefont {Kamionkowski}(2025)}]{Herold:2024nvk}%
  \BibitemOpen
  \bibfield  {author} {\bibinfo {author} {\bibfnamefont {L.}~\bibnamefont {Herold}}\ and\ \bibinfo {author} {\bibfnamefont {M.}~\bibnamefont {Kamionkowski}},\ }\href {https://doi.org/10.1103/PhysRevD.111.083518} {\bibfield  {journal} {\bibinfo  {journal} {Phys. Rev. D}\ }\textbf {\bibinfo {volume} {111}},\ \bibinfo {pages} {083518} (\bibinfo {year} {2025})},\ \Eprint {https://arxiv.org/abs/2412.03546} {arXiv:2412.03546 [astro-ph.CO]} \BibitemShut {NoStop}%
\bibitem [{\citenamefont {Ade}\ \emph {et~al.}(2014{\natexlab{b}})\citenamefont {Ade} \emph {et~al.}}]{Planck:2013jfk}%
  \BibitemOpen
  \bibfield  {author} {\bibinfo {author} {\bibfnamefont {P.~A.~R.}\ \bibnamefont {Ade}} \emph {et~al.} (\bibinfo {collaboration} {Planck}),\ }\href {https://doi.org/10.1051/0004-6361/201321569} {\bibfield  {journal} {\bibinfo  {journal} {Astron. Astrophys.}\ }\textbf {\bibinfo {volume} {571}},\ \bibinfo {pages} {A22} (\bibinfo {year} {2014}{\natexlab{b}})},\ \Eprint {https://arxiv.org/abs/1303.5082} {arXiv:1303.5082 [astro-ph.CO]} \BibitemShut {NoStop}%
\bibitem [{\citenamefont {Blas}\ \emph {et~al.}(2011)\citenamefont {Blas}, \citenamefont {Lesgourgues},\ and\ \citenamefont {Tram}}]{Blas:2011rf}%
  \BibitemOpen
  \bibfield  {author} {\bibinfo {author} {\bibfnamefont {D.}~\bibnamefont {Blas}}, \bibinfo {author} {\bibfnamefont {J.}~\bibnamefont {Lesgourgues}},\ and\ \bibinfo {author} {\bibfnamefont {T.}~\bibnamefont {Tram}},\ }\href {https://doi.org/10.1088/1475-7516/2011/07/034} {\bibfield  {journal} {\bibinfo  {journal} {JCAP}\ }\textbf {\bibinfo {volume} {07}},\ \bibinfo {pages} {034}},\ \Eprint {https://arxiv.org/abs/1104.2933} {arXiv:1104.2933 [astro-ph.CO]} \BibitemShut {NoStop}%
\bibitem [{\citenamefont {Audren}\ \emph {et~al.}(2013)\citenamefont {Audren}, \citenamefont {Lesgourgues}, \citenamefont {Benabed},\ and\ \citenamefont {Prunet}}]{Audren:2012wb}%
  \BibitemOpen
  \bibfield  {author} {\bibinfo {author} {\bibfnamefont {B.}~\bibnamefont {Audren}}, \bibinfo {author} {\bibfnamefont {J.}~\bibnamefont {Lesgourgues}}, \bibinfo {author} {\bibfnamefont {K.}~\bibnamefont {Benabed}},\ and\ \bibinfo {author} {\bibfnamefont {S.}~\bibnamefont {Prunet}},\ }\href {https://doi.org/10.1088/1475-7516/2013/02/001} {\bibfield  {journal} {\bibinfo  {journal} {JCAP}\ }\textbf {\bibinfo {volume} {1302}},\ \bibinfo {pages} {001}},\ \Eprint {https://arxiv.org/abs/1210.7183} {arXiv:1210.7183 [astro-ph.CO]} \BibitemShut {NoStop}%
\bibitem [{\citenamefont {Brinckmann}\ and\ \citenamefont {Lesgourgues}(2018)}]{Brinckmann:2018cvx}%
  \BibitemOpen
  \bibfield  {author} {\bibinfo {author} {\bibfnamefont {T.}~\bibnamefont {Brinckmann}}\ and\ \bibinfo {author} {\bibfnamefont {J.}~\bibnamefont {Lesgourgues}},\ }\href@noop {} {\  (\bibinfo {year} {2018})},\ \Eprint {https://arxiv.org/abs/1804.07261} {arXiv:1804.07261 [astro-ph.CO]} \BibitemShut {NoStop}%
\bibitem [{\citenamefont {Torrado}\ and\ \citenamefont {Lewis}(2021)}]{Torrado:2020dgo}%
  \BibitemOpen
  \bibfield  {author} {\bibinfo {author} {\bibfnamefont {J.}~\bibnamefont {Torrado}}\ and\ \bibinfo {author} {\bibfnamefont {A.}~\bibnamefont {Lewis}},\ }\href {https://doi.org/10.1088/1475-7516/2021/05/057} {\bibfield  {journal} {\bibinfo  {journal} {JCAP}\ }\textbf {\bibinfo {volume} {05}},\ \bibinfo {pages} {057}},\ \Eprint {https://arxiv.org/abs/2005.05290} {arXiv:2005.05290 [astro-ph.IM]} \BibitemShut {NoStop}%
\bibitem [{\citenamefont {Lewis}(2019)}]{Lewis:2019xzd}%
  \BibitemOpen
  \bibfield  {author} {\bibinfo {author} {\bibfnamefont {A.}~\bibnamefont {Lewis}},\ }\href {https://getdist.readthedocs.io} {\  (\bibinfo {year} {2019})},\ \Eprint {https://arxiv.org/abs/1910.13970} {arXiv:1910.13970 [astro-ph.IM]} \BibitemShut {NoStop}%
\bibitem [{\citenamefont {Karwal}\ \emph {et~al.}(2024)\citenamefont {Karwal}, \citenamefont {Patel}, \citenamefont {Bartlett}, \citenamefont {Poulin}, \citenamefont {Smith},\ and\ \citenamefont {Pfeffer}}]{Karwal:2024qpt}%
  \BibitemOpen
  \bibfield  {author} {\bibinfo {author} {\bibfnamefont {T.}~\bibnamefont {Karwal}}, \bibinfo {author} {\bibfnamefont {Y.}~\bibnamefont {Patel}}, \bibinfo {author} {\bibfnamefont {A.}~\bibnamefont {Bartlett}}, \bibinfo {author} {\bibfnamefont {V.}~\bibnamefont {Poulin}}, \bibinfo {author} {\bibfnamefont {T.~L.}\ \bibnamefont {Smith}},\ and\ \bibinfo {author} {\bibfnamefont {D.~N.}\ \bibnamefont {Pfeffer}},\ }\href@noop {} {\  (\bibinfo {year} {2024})},\ \Eprint {https://arxiv.org/abs/2401.14225} {arXiv:2401.14225 [astro-ph.CO]} \BibitemShut {NoStop}%
\bibitem [{\citenamefont {Poulin}\ \emph {et~al.}(2023)\citenamefont {Poulin}, \citenamefont {Smith},\ and\ \citenamefont {Karwal}}]{Poulin:2023lkg}%
  \BibitemOpen
  \bibfield  {author} {\bibinfo {author} {\bibfnamefont {V.}~\bibnamefont {Poulin}}, \bibinfo {author} {\bibfnamefont {T.~L.}\ \bibnamefont {Smith}},\ and\ \bibinfo {author} {\bibfnamefont {T.}~\bibnamefont {Karwal}},\ }\href {https://doi.org/10.1016/j.dark.2023.101348} {\bibfield  {journal} {\bibinfo  {journal} {Phys. Dark Univ.}\ }\textbf {\bibinfo {volume} {42}},\ \bibinfo {pages} {101348} (\bibinfo {year} {2023})},\ \Eprint {https://arxiv.org/abs/2302.09032} {arXiv:2302.09032 [astro-ph.CO]} \BibitemShut {NoStop}%
\bibitem [{\citenamefont {Poudou}\ \emph {et~al.}(2025)\citenamefont {Poudou}, \citenamefont {Simon}, \citenamefont {Montandon}, \citenamefont {Teixeira},\ and\ \citenamefont {Poulin}}]{Poudou:2025qcx}%
  \BibitemOpen
  \bibfield  {author} {\bibinfo {author} {\bibfnamefont {A.}~\bibnamefont {Poudou}}, \bibinfo {author} {\bibfnamefont {T.}~\bibnamefont {Simon}}, \bibinfo {author} {\bibfnamefont {T.}~\bibnamefont {Montandon}}, \bibinfo {author} {\bibfnamefont {E.~M.}\ \bibnamefont {Teixeira}},\ and\ \bibinfo {author} {\bibfnamefont {V.}~\bibnamefont {Poulin}},\ }\href@noop {} {\  (\bibinfo {year} {2025})},\ \Eprint {https://arxiv.org/abs/2503.10485} {arXiv:2503.10485 [astro-ph.CO]} \BibitemShut {NoStop}%
\bibitem [{\citenamefont {Chatrchyan}\ \emph {et~al.}(2025)\citenamefont {Chatrchyan}, \citenamefont {Niedermann}, \citenamefont {Poulin},\ and\ \citenamefont {Sloth}}]{Chatrchyan:2024xjj}%
  \BibitemOpen
  \bibfield  {author} {\bibinfo {author} {\bibfnamefont {A.}~\bibnamefont {Chatrchyan}}, \bibinfo {author} {\bibfnamefont {F.}~\bibnamefont {Niedermann}}, \bibinfo {author} {\bibfnamefont {V.}~\bibnamefont {Poulin}},\ and\ \bibinfo {author} {\bibfnamefont {M.~S.}\ \bibnamefont {Sloth}},\ }\href {https://doi.org/10.1103/PhysRevD.111.043536} {\bibfield  {journal} {\bibinfo  {journal} {Phys. Rev. D}\ }\textbf {\bibinfo {volume} {111}},\ \bibinfo {pages} {043536} (\bibinfo {year} {2025})},\ \Eprint {https://arxiv.org/abs/2408.14537} {arXiv:2408.14537 [astro-ph.CO]} \BibitemShut {NoStop}%
\bibitem [{\citenamefont {Greene}\ and\ \citenamefont {Cyr-Racine}(2024)}]{Greene:2024qis}%
  \BibitemOpen
  \bibfield  {author} {\bibinfo {author} {\bibfnamefont {K.}~\bibnamefont {Greene}}\ and\ \bibinfo {author} {\bibfnamefont {F.-Y.}\ \bibnamefont {Cyr-Racine}},\ }\href {https://doi.org/10.1103/PhysRevD.110.043524} {\bibfield  {journal} {\bibinfo  {journal} {Phys. Rev. D}\ }\textbf {\bibinfo {volume} {110}},\ \bibinfo {pages} {043524} (\bibinfo {year} {2024})},\ \Eprint {https://arxiv.org/abs/2403.05619} {arXiv:2403.05619 [astro-ph.CO]} \BibitemShut {NoStop}%
\bibitem [{\citenamefont {Efstathiou}\ \emph {et~al.}(2024)\citenamefont {Efstathiou}, \citenamefont {Rosenberg},\ and\ \citenamefont {Poulin}}]{Efstathiou:2023fbn}%
  \BibitemOpen
  \bibfield  {author} {\bibinfo {author} {\bibfnamefont {G.}~\bibnamefont {Efstathiou}}, \bibinfo {author} {\bibfnamefont {E.}~\bibnamefont {Rosenberg}},\ and\ \bibinfo {author} {\bibfnamefont {V.}~\bibnamefont {Poulin}},\ }\href {https://doi.org/10.1103/PhysRevLett.132.221002} {\bibfield  {journal} {\bibinfo  {journal} {Phys. Rev. Lett.}\ }\textbf {\bibinfo {volume} {132}},\ \bibinfo {pages} {221002} (\bibinfo {year} {2024})},\ \Eprint {https://arxiv.org/abs/2311.00524} {arXiv:2311.00524 [astro-ph.CO]} \BibitemShut {NoStop}%
\bibitem [{\citenamefont {Hu}\ \emph {et~al.}(2001)\citenamefont {Hu}, \citenamefont {Fukugita}, \citenamefont {Zaldarriaga},\ and\ \citenamefont {Tegmark}}]{Hu:2000ti}%
  \BibitemOpen
  \bibfield  {author} {\bibinfo {author} {\bibfnamefont {W.}~\bibnamefont {Hu}}, \bibinfo {author} {\bibfnamefont {M.}~\bibnamefont {Fukugita}}, \bibinfo {author} {\bibfnamefont {M.}~\bibnamefont {Zaldarriaga}},\ and\ \bibinfo {author} {\bibfnamefont {M.}~\bibnamefont {Tegmark}},\ }\href {https://doi.org/10.1086/319449} {\bibfield  {journal} {\bibinfo  {journal} {Astrophys. J.}\ }\textbf {\bibinfo {volume} {549}},\ \bibinfo {pages} {669} (\bibinfo {year} {2001})},\ \Eprint {https://arxiv.org/abs/astro-ph/0006436} {arXiv:astro-ph/0006436} \BibitemShut {NoStop}%
\bibitem [{\citenamefont {Raveri}\ and\ \citenamefont {Hu}(2019)}]{Raveri:2018wln}%
  \BibitemOpen
  \bibfield  {author} {\bibinfo {author} {\bibfnamefont {M.}~\bibnamefont {Raveri}}\ and\ \bibinfo {author} {\bibfnamefont {W.}~\bibnamefont {Hu}},\ }\href {https://doi.org/10.1103/PhysRevD.99.043506} {\bibfield  {journal} {\bibinfo  {journal} {Phys. Rev. D}\ }\textbf {\bibinfo {volume} {99}},\ \bibinfo {pages} {043506} (\bibinfo {year} {2019})},\ \Eprint {https://arxiv.org/abs/1806.04649} {arXiv:1806.04649 [astro-ph.CO]} \BibitemShut {NoStop}%
\bibitem [{\citenamefont {Mortonson}\ and\ \citenamefont {Hu}(2009)}]{Mortonson:2009xk}%
  \BibitemOpen
  \bibfield  {author} {\bibinfo {author} {\bibfnamefont {M.~J.}\ \bibnamefont {Mortonson}}\ and\ \bibinfo {author} {\bibfnamefont {W.}~\bibnamefont {Hu}},\ }\href {https://doi.org/10.1103/PhysRevD.80.027301} {\bibfield  {journal} {\bibinfo  {journal} {Phys. Rev. D}\ }\textbf {\bibinfo {volume} {80}},\ \bibinfo {pages} {027301} (\bibinfo {year} {2009})},\ \Eprint {https://arxiv.org/abs/0906.3016} {arXiv:0906.3016 [astro-ph.CO]} \BibitemShut {NoStop}%
\bibitem [{\citenamefont {Lodha}\ \emph {et~al.}(2025)\citenamefont {Lodha} \emph {et~al.}}]{DESI:2025fii}%
  \BibitemOpen
  \bibfield  {author} {\bibinfo {author} {\bibfnamefont {K.}~\bibnamefont {Lodha}} \emph {et~al.} (\bibinfo {collaboration} {DESI}),\ }\href@noop {} {\  (\bibinfo {year} {2025})},\ \Eprint {https://arxiv.org/abs/2503.14743} {arXiv:2503.14743 [astro-ph.CO]} \BibitemShut {NoStop}%
\bibitem [{\citenamefont {Harrington}\ \emph {et~al.}(2016)\citenamefont {Harrington} \emph {et~al.}}]{Harrington:2016jrz}%
  \BibitemOpen
  \bibfield  {author} {\bibinfo {author} {\bibfnamefont {K.}~\bibnamefont {Harrington}} \emph {et~al.},\ }\href {https://doi.org/10.1117/12.2233125} {\bibfield  {journal} {\bibinfo  {journal} {Proc. SPIE Int. Soc. Opt. Eng.}\ }\textbf {\bibinfo {volume} {9914}},\ \bibinfo {pages} {99141K} (\bibinfo {year} {2016})},\ \Eprint {https://arxiv.org/abs/1608.08234} {arXiv:1608.08234 [astro-ph.IM]} \BibitemShut {NoStop}%
\bibitem [{\citenamefont {Watts}\ \emph {et~al.}(2018)\citenamefont {Watts} \emph {et~al.}}]{Watts:2018etg}%
  \BibitemOpen
  \bibfield  {author} {\bibinfo {author} {\bibfnamefont {D.~J.}\ \bibnamefont {Watts}} \emph {et~al.},\ }\href {https://doi.org/10.3847/1538-4357/aad283} {\bibfield  {journal} {\bibinfo  {journal} {Astrophys. J.}\ }\textbf {\bibinfo {volume} {863}},\ \bibinfo {pages} {121} (\bibinfo {year} {2018})},\ \Eprint {https://arxiv.org/abs/1801.01481} {arXiv:1801.01481 [astro-ph.CO]} \BibitemShut {NoStop}%
\bibitem [{\citenamefont {Li}\ \emph {et~al.}(2025)\citenamefont {Li} \emph {et~al.}}]{CLASS:2025khf}%
  \BibitemOpen
  \bibfield  {author} {\bibinfo {author} {\bibfnamefont {Y.}~\bibnamefont {Li}} \emph {et~al.} (\bibinfo {collaboration} {CLASS}),\ }\href@noop {} {\  (\bibinfo {year} {2025})},\ \Eprint {https://arxiv.org/abs/2501.11904} {arXiv:2501.11904 [astro-ph.CO]} \BibitemShut {NoStop}%
\bibitem [{\citenamefont {Hazumi}\ \emph {et~al.}(2020)\citenamefont {Hazumi} \emph {et~al.}}]{LiteBIRD:2020khw}%
  \BibitemOpen
  \bibfield  {author} {\bibinfo {author} {\bibfnamefont {M.}~\bibnamefont {Hazumi}} \emph {et~al.} (\bibinfo {collaboration} {LiteBIRD}),\ }\href {https://doi.org/10.1117/12.2563050} {\bibfield  {journal} {\bibinfo  {journal} {Proc. SPIE Int. Soc. Opt. Eng.}\ }\textbf {\bibinfo {volume} {11443}},\ \bibinfo {pages} {114432F} (\bibinfo {year} {2020})},\ \Eprint {https://arxiv.org/abs/2101.12449} {arXiv:2101.12449 [astro-ph.IM]} \BibitemShut {NoStop}%
\bibitem [{\citenamefont {Sakamoto}\ \emph {et~al.}(2022)\citenamefont {Sakamoto}, \citenamefont {Ahn}, \citenamefont {Ichiki}, \citenamefont {Moon},\ and\ \citenamefont {Hasegawa}}]{Sakamoto:2022nth}%
  \BibitemOpen
  \bibfield  {author} {\bibinfo {author} {\bibfnamefont {H.}~\bibnamefont {Sakamoto}}, \bibinfo {author} {\bibfnamefont {K.}~\bibnamefont {Ahn}}, \bibinfo {author} {\bibfnamefont {K.}~\bibnamefont {Ichiki}}, \bibinfo {author} {\bibfnamefont {H.}~\bibnamefont {Moon}},\ and\ \bibinfo {author} {\bibfnamefont {K.}~\bibnamefont {Hasegawa}},\ }\href {https://doi.org/10.3847/1538-4357/ac6668} {\bibfield  {journal} {\bibinfo  {journal} {Astrophys. J.}\ }\textbf {\bibinfo {volume} {930}},\ \bibinfo {pages} {140} (\bibinfo {year} {2022})},\ \Eprint {https://arxiv.org/abs/2202.04263} {arXiv:2202.04263 [astro-ph.CO]} \BibitemShut {NoStop}%
\bibitem [{\citenamefont {Liu}\ \emph {et~al.}(2016)\citenamefont {Liu}, \citenamefont {Pritchard}, \citenamefont {Allison}, \citenamefont {Parsons}, \citenamefont {Seljak},\ and\ \citenamefont {Sherwin}}]{Liu:2015txa}%
  \BibitemOpen
  \bibfield  {author} {\bibinfo {author} {\bibfnamefont {A.}~\bibnamefont {Liu}}, \bibinfo {author} {\bibfnamefont {J.~R.}\ \bibnamefont {Pritchard}}, \bibinfo {author} {\bibfnamefont {R.}~\bibnamefont {Allison}}, \bibinfo {author} {\bibfnamefont {A.~R.}\ \bibnamefont {Parsons}}, \bibinfo {author} {\bibfnamefont {U.}~\bibnamefont {Seljak}},\ and\ \bibinfo {author} {\bibfnamefont {B.~D.}\ \bibnamefont {Sherwin}},\ }\href {https://doi.org/10.1103/PhysRevD.93.043013} {\bibfield  {journal} {\bibinfo  {journal} {Phys. Rev. D}\ }\textbf {\bibinfo {volume} {93}},\ \bibinfo {pages} {043013} (\bibinfo {year} {2016})},\ \Eprint {https://arxiv.org/abs/1509.08463} {arXiv:1509.08463 [astro-ph.CO]} \BibitemShut {NoStop}%
\bibitem [{\citenamefont {Shmueli}\ \emph {et~al.}(2023)\citenamefont {Shmueli}, \citenamefont {Sarkar},\ and\ \citenamefont {Kovetz}}]{Shmueli:2023box}%
  \BibitemOpen
  \bibfield  {author} {\bibinfo {author} {\bibfnamefont {G.}~\bibnamefont {Shmueli}}, \bibinfo {author} {\bibfnamefont {D.}~\bibnamefont {Sarkar}},\ and\ \bibinfo {author} {\bibfnamefont {E.~D.}\ \bibnamefont {Kovetz}},\ }\href {https://doi.org/10.1103/PhysRevD.108.083531} {\bibfield  {journal} {\bibinfo  {journal} {Phys. Rev. D}\ }\textbf {\bibinfo {volume} {108}},\ \bibinfo {pages} {083531} (\bibinfo {year} {2023})},\ \Eprint {https://arxiv.org/abs/2305.07056} {arXiv:2305.07056 [astro-ph.CO]} \BibitemShut {NoStop}%
\bibitem [{\citenamefont {{Finkelstein}}\ \emph {et~al.}(2024)\citenamefont {{Finkelstein}} \emph {et~al.}}]{2024ApJ...969L...2F}%
  \BibitemOpen
  \bibfield  {author} {\bibinfo {author} {\bibfnamefont {S.~L.}\ \bibnamefont {{Finkelstein}}} \emph {et~al.},\ }\href {https://doi.org/10.3847/2041-8213/ad4495} {\bibfield  {journal} {\bibinfo  {journal} {Astrophys. J. Lett.}\ }\textbf {\bibinfo {volume} {969}},\ \bibinfo {eid} {L2} (\bibinfo {year} {2024})},\ \Eprint {https://arxiv.org/abs/2311.04279} {arXiv:2311.04279 [astro-ph.GA]} \BibitemShut {NoStop}%
\bibitem [{\citenamefont {{Castellano}}\ \emph {et~al.}(2022)\citenamefont {{Castellano}} \emph {et~al.}}]{2022ApJ...938L..15C}%
  \BibitemOpen
  \bibfield  {author} {\bibinfo {author} {\bibfnamefont {M.}~\bibnamefont {{Castellano}}} \emph {et~al.},\ }\href {https://doi.org/10.3847/2041-8213/ac94d0} {\bibfield  {journal} {\bibinfo  {journal} {Astrophys. J. Lett.}\ }\textbf {\bibinfo {volume} {938}},\ \bibinfo {eid} {L15} (\bibinfo {year} {2022})},\ \Eprint {https://arxiv.org/abs/2207.09436} {arXiv:2207.09436 [astro-ph.GA]} \BibitemShut {NoStop}%
\bibitem [{\citenamefont {{Eisenstein}}\ \emph {et~al.}(2023)\citenamefont {{Eisenstein}} \emph {et~al.}}]{2023arXiv230602465E}%
  \BibitemOpen
  \bibfield  {author} {\bibinfo {author} {\bibfnamefont {D.~J.}\ \bibnamefont {{Eisenstein}}} \emph {et~al.},\ }\href {https://doi.org/10.48550/arXiv.2306.02465} {\bibfield  {journal} {\bibinfo  {journal} {arXiv e-prints}\ ,\ \bibinfo {eid} {arXiv:2306.02465}} (\bibinfo {year} {2023})},\ \Eprint {https://arxiv.org/abs/2306.02465} {arXiv:2306.02465 [astro-ph.GA]} \BibitemShut {NoStop}%
\bibitem [{\citenamefont {{Harikane}}\ \emph {et~al.}(2023)\citenamefont {{Harikane}} \emph {et~al.}}]{2023ApJS..265....5H}%
  \BibitemOpen
  \bibfield  {author} {\bibinfo {author} {\bibfnamefont {Y.}~\bibnamefont {{Harikane}}} \emph {et~al.},\ }\href {https://doi.org/10.3847/1538-4365/acaaa9} {\bibfield  {journal} {\bibinfo  {journal} {Astrophys. J. Supp.}\ }\textbf {\bibinfo {volume} {265}},\ \bibinfo {eid} {5} (\bibinfo {year} {2023})},\ \Eprint {https://arxiv.org/abs/2208.01612} {arXiv:2208.01612 [astro-ph.GA]} \BibitemShut {NoStop}%
\bibitem [{\citenamefont {Mu\~noz}\ \emph {et~al.}(2024)\citenamefont {Mu\~noz}, \citenamefont {Mirocha}, \citenamefont {Chisholm}, \citenamefont {Furlanetto},\ and\ \citenamefont {Mason}}]{Munoz:2024fas}%
  \BibitemOpen
  \bibfield  {author} {\bibinfo {author} {\bibfnamefont {J.~B.}\ \bibnamefont {Mu\~noz}}, \bibinfo {author} {\bibfnamefont {J.}~\bibnamefont {Mirocha}}, \bibinfo {author} {\bibfnamefont {J.}~\bibnamefont {Chisholm}}, \bibinfo {author} {\bibfnamefont {S.~R.}\ \bibnamefont {Furlanetto}},\ and\ \bibinfo {author} {\bibfnamefont {C.}~\bibnamefont {Mason}},\ }\href {https://doi.org/10.1093/mnrasl/slae086} {\bibfield  {journal} {\bibinfo  {journal} {Mon. Not. Roy. Astron. Soc.}\ }\textbf {\bibinfo {volume} {535}},\ \bibinfo {pages} {L37} (\bibinfo {year} {2024})},\ \Eprint {https://arxiv.org/abs/2404.07250} {arXiv:2404.07250 [astro-ph.CO]} \BibitemShut {NoStop}%
\bibitem [{\citenamefont {Sailer}\ \emph {et~al.}(2025)\citenamefont {Sailer}, \citenamefont {Farren}, \citenamefont {Ferraro},\ and\ \citenamefont {White}}]{Sailer:2025lxj}%
  \BibitemOpen
  \bibfield  {author} {\bibinfo {author} {\bibfnamefont {N.}~\bibnamefont {Sailer}}, \bibinfo {author} {\bibfnamefont {G.~S.}\ \bibnamefont {Farren}}, \bibinfo {author} {\bibfnamefont {S.}~\bibnamefont {Ferraro}},\ and\ \bibinfo {author} {\bibfnamefont {M.}~\bibnamefont {White}},\ }\href@noop {} {\  (\bibinfo {year} {2025})},\ \Eprint {https://arxiv.org/abs/2504.16932} {arXiv:2504.16932 [astro-ph.CO]} \BibitemShut {NoStop}%
\end{thebibliography}%

\end{document}